\documentclass[useAMS,usenatbib]{mn2e}
\usepackage{graphicx}
\usepackage{hyperref}
\usepackage{natbib}
\usepackage{subfig}
\usepackage{epsfig} 
\usepackage{ifpdf} 
\usepackage{fixltx2e}
\usepackage{txfonts}
\usepackage{xcolor}
\usepackage[all]{hypcap}

\newcommand{\be}{\begin{equation}}                                 
\newcommand{\ee}{\end{equation}}                                   
\newcommand{\bea}{\begin{eqnarray}}                                
\newcommand{\eea}{\end{eqnarray}}                                  
                                                 
\title[Thin-disc instability]{Three-dimensional, global, radiative GRMHD simulations of a thermally unstable disc}
\author[B. Mishra, P. C. Fragile, C. L. Johnson, W. Klu\'zniak]{B. Mishra$^{1}$\thanks{E-mail:mbhupe@camk.edu.pl}, P. C. Fragile$^{2}$\thanks{E-mail:FragileP@cofc.edu},  C. L. Johnson$^{2}$, W. Klu\'zniak$^{1}$\thanks{E-mail:wlodek@camk.edu.pl}\\
$^{1}$Nicolaus Copernicus Astronomical Center, ul. Bartycka 18, Warsaw, 00-716, Poland\\
$^{2}$Department of Physics and Astronomy, College of Charleston, 66 George Street, Charleston SC 29424, USA}
\begin{document}
\date{Accepted *** Received ***}
\pagerange{\pageref{firstpage}--\pageref{lastpage}} \pubyear{2016}
\maketitle
\label{firstpage}
\begin{abstract}
We present results of a set of three-dimensional, general relativistic radiation magnetohydrodynamics simulations of thin accretion discs around a non-rotating black hole to test their thermal stability.  We consider two cases, one that is initially radiation-pressure dominated and expected to be thermally unstable and another that is initially gas-pressure dominated and expected to remain stable.  Indeed, we find that cooling dominates over heating in the radiation-pressure-dominated model, causing the disc to collapse vertically on roughly the local cooling timescale.  We also find that heating and cooling within the disc have a different dependence on the mid-plane pressure---a prerequisite of thermal instability.  Comparison of our data with the relevant thin-disc thermal equilibrium curve suggests that our disc may be headed for the thermally stable, gas-pressure-dominated branch.  However, because the disc collapses to the point that we are no longer able to resolve it, we had to terminate the simulation.  On the other hand, the gas-pressure-dominated model, which was run for twice as long as the radiation-pressure-dominated one, remains stable, with heating and cooling roughly in balance. Finally, the radiation pressure dominated simulation shows some evidence of viscous instability. The strongest evidence is in plots of surface density, which show the disc breaking up into rings.
\end{abstract}

\section{Introduction}
It has been over forty years since the seminal paper on geometrically thin accretion discs was published by \citet{ss1973}. This model prescribed three different regions in such discs: a radiation-pressure-dominated (inner) region, with the opacity dominated by electron scattering; a gas-pressure-dominated (middle) region, with the dominant opacity again due to electron scattering; and a gas-pressure-dominated (outer) region, with opacity dominated by free-free absorption. 

Linear stability analysis of the radiation-pressure-dominated region indicated that it should be unstable \citep{ss1976}. The origin of this instability is the assumption that the anomalous stress, $\tau_{r\phi}$, that drives accretion is proportional to the total (gas plus radiation) pressure. This actually leads to two instabilities, one being thermal \citep{ss1976} and the other viscous \citep{lightman1974}. In this work we are mainly focused on the thermal instability. \citet{pringle1976} and \citet{piran1978} showed that the thermal instability originates in the different dependence of the heating and cooling rates per unit area on the disc mid-plane temperature, for a constant surface density, $\Sigma$. The ratio of heating rate per unit area to cooling rate per unit area is proportional to the fourth power of the mid-plane temperature \citep{pringle1976}.  A small fluctuation of temperature in an equilibrium disc can lead to excess heating resulting in an expanding disc, or excess cooling can lead to a collapse of the disc, as in recent numerical simulations \citep{jiang2013,sadowski16}.  All of this assumes that any magnetic fields in the disc are weak.  If strong magnetic fields are present, they can, in principle, stabilize the disc \citep{begelman07,oda09,sadowski16}.     
 
In recent years, local, shearing-box \citep{bran1995,stone1996} numerical simulations have been performed to study the thermal stability of radiation-pressure-dominated discs \citep{turner2004,hirose2009,jiang2013}. 
The first radiation MHD simulation using a stratified shearing box was performed by \citet{turner2004}. In this simulation, even with a radiation-to-gas pressure ratio of $\sim 14$, the disc did not show any thermal instability. However, those results are suspect, as the photosphere was not captured within the simulation domain (during the expansion phase caused by heating). Furthermore, half of the mass was lost from the boundaries of the simulation box and $27\%$ of work done on the box disappeared due to numerical losses. \citet{hirose2009} repeated the radiation-pressure-dominated shearing box simulations with a better energy conservation scheme and larger box to retain both (top and bottom) photospheres within the simulation domain. They also greatly reduced the mass loss through the box boundaries. These simulations, too, showed no thermal instability. The analytic and numerical results were thus in conflict until new shearing box simulations were performed by \citet{jiang2013}.  Depending on the central density and the ratio of radiation pressure to gas pressure, \citet{jiang2013} found all of their discs to either expand or collapse on a timescale of tens of orbits.  They were also able to demonstrate that the previous contradictory results were owing to the use of too small boxes. 

We expand the previous shearing box results into the domain of global simulations using the {\it Cosmos++} general relativistic radiation magnetohydrodynamic (GRRMHD) code \citep{anninos2005,fragile2012,fragile2014}.  We compare two general cases, one gas-pressure-dominated, the other radiation-pressure-dominated.  The radiation-pressure-dominated case has parameters chosen to closely match those of one of the unstable shearing-box cases of \citet{jiang2013}.  

In addition to investigating thermal instability, global simulations also open up the possibility of testing the viscous stability of discs.  This cannot be done with shearing box simulations, as by design they maintain a constant surface density, $\Sigma$, while the viscous instability induces radial variations in $\Sigma$.

Before proceeding to describe our simulations and results, let us mention a few points about notation: we use standard index notation where repeated indices imply summation, Greek indices cover the four spacetime dimensions, and Latin indices cover the three spatial dimensions. The metric signature is taken to be -+++. Additionally, most equations are presented in units where $G=c=1$.

\section{Numerical setup}
\subsection{Initial configurations}

In order to make connections with previous shearing box \citep[esp.][]{jiang2013} and global, pseudo-Newtonian thin-disc \citep{reynolds2009} simulations, instead of starting from the Shakura-Sunyaev solution, we initialized a slab of gas of uniform thickness, orbiting everywhere at the local Keplerian frequency. We chose a black hole mass of $M_\mathrm{BH} = 6.62 M_\odot$, for close comparison with \citet{jiang2013}.  The initial density profile of the azimuthally symmetric disc is 
\begin{equation}
\rho(R,z) = \frac{\rho_0e^{-z^2/2H^2}} {1 + e^{(R_i -R)/H}}~,
\label{initial_density}
\end{equation}
where $\rho_0$ is the mid-plane density, $H$ is the initial height of the disc, and $R_i$ is the inner radius of the initial disc. We fix the initial inner radius to that of the innermost stable circular orbit (ISCO) and use an exponential cutoff to smooth the transition there.  The initial disc structure is thus entirely governed by the chosen mid-plane density and disc height. We consider two cases, one with mid-plane density $\rho_0 = 10^{-3}~\mathrm{g}\,\mathrm{cm}^{-3}$ and height $H_0 = 0.4\,r_g$, where $r_g = GM/c^2$ is the gravitational radius, and the other with $\rho_0 = 10^{-6}~\mathrm{g}\,\mathrm{cm}^{-3}$ and $H_0 = 0.3\,r_g$. The first case matches the RSVET model in the shearing-box simulations of \citet{jiang2013}. The \citet{jiang2013} simulations were done for a section of disc centered at $R=30\,r_g$, whereas we are performing global simulations, so the correspondence is imperfect. In our case, the disc is initially radiation pressure dominated, with $10 \lesssim P_{\mathrm{rad}}/P_{\mathrm{gas}} \lesssim 1000$. We refer to this setup as RADPHR or RADPLR depending on the resolution used (see Table~\ref{models}). 

For RADPHR and RADPLR, we take the gas and radiation to be initially in local thermodynamic equilibrium (LTE), which means the total pressure is distributed between the two (magnetic pressure is initially negligible), as
\begin{equation}
P_\mathrm{tot} = P_\mathrm{gas} + P_\mathrm{rad}~,
\label{eq:ptot}
\end{equation}
where the total pressure, $P_\mathrm{tot}$, is initially taken to be
\begin{equation}
P_\mathrm{tot} = \frac{G M_\mathrm{BH} H^2}{R(R-2\,r_g)^2}\rho(R,z)~,
\label{eq:PW}
\end{equation}
as in \citet{reynolds2009}. The disc is thus initially in approximate hydrostatic equilibrium: on the one hand, the pressure value given by equation (\ref{eq:PW}) exceeds the value required for vertical hydrostatic equilibrium in the Schwarzschild metric by a factor of $\approx 1.3$; on the other, until MRI develops, there is no heating in the disc.  Substituting the expressions for radiation pressure (in local thermodynamic equilibrium) and ideal gas pressure into equation (\ref{eq:ptot}), we have
\begin{equation}
\frac{1}{3}a_\mathrm{R} T_\mathrm{gas}^4 + \frac{k_\mathrm{b}\rho T_\mathrm{gas}}{\mu} - P_\mathrm{tot} = 0~,
\label{lte}
\end{equation}
where $a_\mathrm{R} = 4\sigma/c$ is the radiation constant. This is a quartic equation for the gas temperature with four possible roots, though only one is positive and real. 

\begin{figure}
\centering
\includegraphics[width=1\columnwidth]{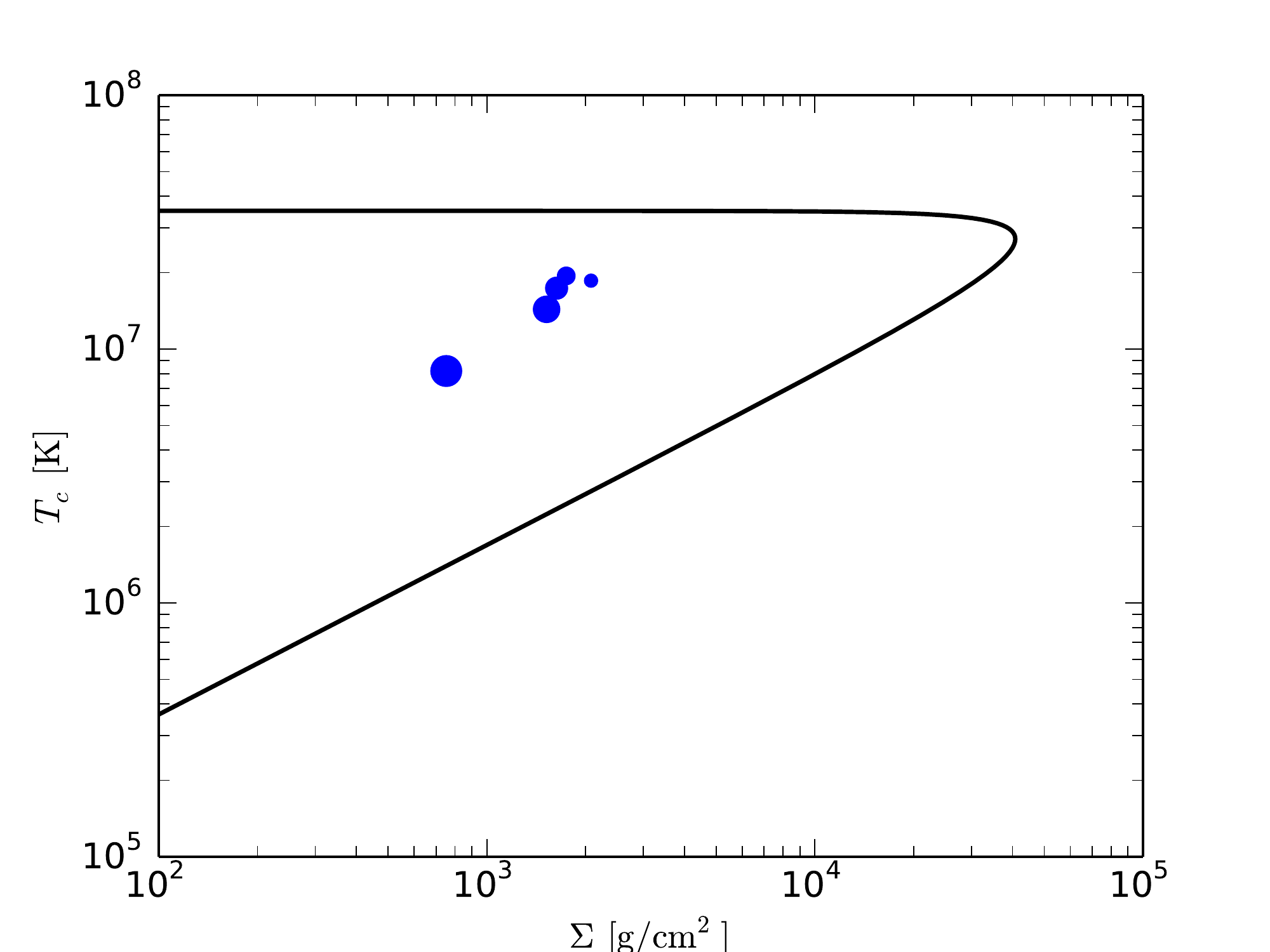}
\caption{Thermal equilibrium ($T_c$--$\Sigma$) diagram for a thin-disc solution at $R = 10\,r_g$.  The solid line is for a standard thin disc (excluding advection) with $\alpha = 0.02$ (see Fig.~\ref{alpha_radphr}).  The blue points show the evolution of the RADPHR simulation (increasing point sizes correspond to $t = \{0,1108,2215,3323,4430\}~GM/c^3$, respectively).}
\label{scurve}
\end{figure}

For the chosen values of the initial density and disc height, these simulations do not begin in thermal equilibrium.  Fig. \ref{scurve} shows a thermal equilibrium curve at $R = 10\,r_g$ for an $\alpha = 0.02$ thin disc.  Our simulations start between the unstable, radiation-pressure-dominated and stable, gas-pressure-dominated branches.

The second case we consider is somewhat different in terms of its thermodynamic properties. This lower density disc is initially gas pressure dominated, with $10^{-11}\lesssim P_{\mathrm{rad}}/P_{\mathrm{gas}} \lesssim 10^{-7}$ for entire disc. We refer to this setup as GASPLR or GASPHR, again depending on the resolution (Table~\ref{models}). GASPHR and GASPLR have an initial disc thickness $H_0 = 0.3 r_g$. In this case, we do not assume LTE, initially. Instead, we define a uniform, initial radiation temperature, $T_{\mathrm{rad}} = 10^4$ K, considerably below the initial gas temperature.  We do this to ensure that the simulation starts out completely gas-pressure-dominated.  In effect, we are artificially suppressing the initial radiation field.  However, the surface density in this case is so low ($\le 0.735~\mathrm{g~cm^{-2}}$) that the disc is effectively optically thin to electron scattering, so the coupling between the gas and radiation is weak.

\subsection{Radiation fields}

The $\mathbf{M}_1$ closure scheme in {\it Cosmos++} \citep{fragile2014} evolves the radiation energy density in the radiation rest frame, $E_R$, and the spatial components of the radiation rest frame 4-velocity, $u^i_R$ (see Section \ref{sec:scheme}). However, it is easier to initialize our simulations by defining the radiation fields in the fluid frame.  We already described in the previous section how we determine the initial radiation temperature, $T_\mathrm{rad}$.  This temperature can be used to define the radiation energy density in the fluid frame 
\begin{equation}
E_\mathrm{rad} = a_\mathrm{R}T_\mathrm{rad}^4~.
\label{erad}
\end{equation}
The initial radiative flux, $F^i$, is then calculated from the gradient of $E_\mathrm{rad}$. With these quantities, we can construct the contravariant radiation stress-energy tensor 
\begin{equation}
R^{\alpha\beta} = E_\mathrm{rad}u^\alpha u^\beta + F^\alpha u^\beta + F^\beta u ^\alpha + \frac{1}{3}E_\mathrm{rad}h^{\alpha\beta}~,
\label{Rab}
\end{equation}
where $h^{\alpha\beta} = g^{\alpha\beta} + u^\alpha u^\beta$ is the projection tensor and $u^\alpha$ is the fluid 4-velocity.

We can equate the radiation stress-energy tensor from equation (\ref{Rab}) to the form written in terms of the radiation rest frame variables:
\begin{equation}
R^{\alpha\beta} = \frac{4}{3}E_R u_R^\alpha u_R^\beta + \frac{1}{3}E_Rg^{\alpha\beta}~.
\label{radTensor}
\end{equation}
Following \citet{2013MNRAS.429.3533S} and \citet{fragile2014}, we start with the following two relationships, both of which come from equation (\ref{radTensor}):
\begin{equation}
g_{\alpha\beta}R^{t\alpha}R^{t\beta} = -\frac{8}{9}E_\mathrm{R}^2(u^t_\mathrm{R})^2 + \frac{1}{9}E_\mathrm{R}^2g^{tt}
\label{gmunurmunu}
\end{equation}
and
\begin{equation}
R^{tt} = \frac{4}{3}E_\mathrm{R}(u^t_\mathrm{R})^2 + \frac{1}{3}E_\mathrm{R}g^{tt}~.
\label{rtt}
\end{equation}
Using these, we solve for the radiation energy density in the radiation rest frame, $E_R$, and the time component of the radiation rest frame four-velocity, $u^t_{R}$. With these, the remaining spatial components of the radiation rest frame 4-velocity, $u^i_R$, can easily be obtained from equation (\ref{radTensor}).

\subsection{Magnetic field setup}

To seed the magneto-rotational instability (MRI) inside the disc, we impose a weak magnetic field ($\beta = P_\mathrm{tot}/P_\mathrm{mag} \geq 10$) on top of our hydrodynamical setup.  The MRI is necessary to drive the accretion of matter into the black hole and transport angular momentum outwards \citep{balbus1991}.  The turbulence of the MRI will also heat the disc, which is important to our goal of studying thermal stability. 

It is important for the initialized magnetic field to be divergence free. The easiest way to accomplish this is to initialize the magnetic field starting from the vector potential. For our thin discs, we set $A_r = A_z = 0$ and
\begin{equation}
A_\phi = \frac{\sqrt{P_\mathrm{gas}}\sin\left(2\pi R/5H\right)}{1 + e^\Delta}~,
\label{scalrPot}
\end{equation}
where
\begin{equation}
\Delta = 10\left\{ \frac{z^2}{H^2} + \left(\frac{H}{R - R_\mathrm{ISCO}}\right)^2 - 1 \right\}~
\label{delta}
\end{equation}
and $R_\mathrm{ISCO}=6\,r_g$ is the radius of the ISCO. The effect is to create small magnetic field loops of roughly the same size as the disc thickness and alternating polarity centered along the mid-plane of the disc.  The generalized curl of this vector potential
\begin{equation}
\mathcal{B}^i = \epsilon^{ij\phi} \partial_j A_\phi
\end{equation}
then gives the appropriate magnetic field components. The strength of the magnetic field is scaled to match the chosen $\beta$. In order to keep the magnetic field divergence free during the evolution, we use the staggered, {\it constrained transport} scheme described in \citet{fragile2012}. 

\subsection{Grid parameters}

In thin-disc simulations, a big challenge is to have enough resolution to capture the MRI. To help with this, we adopt a cylindrical, Kerr-Schild coordinate system computed by transformation from the usual spherical Kerr-Schild coordinate system ($R = r\sin\theta$ and $z = r\cos\theta$).  To further improve the resolution near the disc mid-plane, we space the $n_z = 160$ zones using a logarithmic coordinate 
\begin{equation}
x_3 = \pm \ln \left( \frac{n_z \vert z \vert - L_z}{L_z} \right)~,
\end{equation}
where $L_z = 20 H$ is the total box height and the sign of the expression comes from the sign of $z$.  We also employ a logarithmic grid in the radial direction with $n_R = 192$ zones spaced as
\begin{equation}
x_1 = 1 + \ln \left(\frac{R}{r_\mathrm{BH}} \right)~,
\end{equation}
where $r_\mathrm{BH}=2\,r_g$ is the black hole radius. The radial domain runs over $4 r_g \le R \le 40 r_g$. A linear spacing is used along the azimuthal direction, $\phi$, with $n_\phi = 32$ (low resolution) or $n_\phi = 64$ (high resolution). To reduce the computational expense we only simulate the $0 \le \phi \le 0.5\,\pi$ wedge in the azimuthal direction. 

We use outflow boundary conditions in the inner radial and top and bottom vertical boundaries, which means all the gas variables except the normal component of the flow velocity are copied from the last active zones to the ghost zones. If the normal velocity component points outward, then it, too, is copied to the ghost zone.  Otherwise, it is adjusted such that the normal component of the velocity is zero at the boundary face.  The outer radial boundary uses a constant boundary condition, which means that all variables will retain their initial values in the outer radial ghost zones. The azimuthal boundary conditions are periodic.

\begin{table}[]
\caption{Simulation parameters for the four models considered. The model names reflect whether the simulation is radiation- (RADP) or gas- (GASP) pressure dominated and high (HR) or low (LR) resolution. The number of cells along the radial, $n_\mathrm{R} = 192$, and vertical, $n_\mathrm{z} = 160$, directions are fixed for all four cases.}
\centering                          
\begin{tabular*}{0.5\textwidth}{@{\extracolsep{\fill}}  l   c c c c c c}
\\
\hline \hline
	Model & $\rho_0$  & $P_\mathrm{rad}/P_\mathrm{gas}$ & $192\times n_\phi\times 160$ &Run Time
\\
	 & $(\mathrm{g\,cm^{-3}})$  &  &  & $(t_\mathrm{ISCO} = 92.3 GM/c^3)$& 
\\
\hline \hline
\\
RADPHR    &  $10^{-3}$ &200 & $64$ &48\\
RADPLR      & $10^{-3}$ &200 &$32$ &42\\
GASPHR &$10^{-6}$ & $10^{-7}$ &$64$ &32\\
GASPLR &$10^{-6}$ & $10^{-7}$ &$32$ &80\\
\hline        \hline                                   
\end{tabular*}
\label{models}
\end{table}

\subsection{Radiative GRMHD scheme}
\label{sec:scheme}

We solve the coupled radiative magnetohydrodynamics equations using the \textbf{$\textrm{M}_1$} closure scheme to handle both the optically thick and thin limits, as described in \citet{fragile2014}. Along with the radiation stress-energy tensor, defined in equation (\ref{radTensor}), we need the MHD stress-energy tensor
\begin{equation}
T^{\alpha\beta} = \left(\rho + \rho\varepsilon + P_{\mathrm{gas}} + b^2\right)u^{\alpha}u^{\beta} + \left(P_{\mathrm{gas}} + P_\mathrm{mag}\right)g^{\alpha\beta} - b^\alpha b^\beta,
\end{equation}
where $\rho$ is rest mass density, $\varepsilon$ is specific internal energy, $P_{\mathrm{gas}}$ is gas pressure, defined using the ideal gas equation of state, $P_{\mathrm{gas}} = (\gamma -1)\rho\varepsilon$, with $\gamma =5/3$, and $b^\alpha$ is the contravariant magnetic 4-field, measured by an observer co-moving with fluid.

We aim to solve the following set of conservation equations for the mass
\begin{equation}
\left(\rho u^{\beta}\right)_{;\,\beta} = 0,
\end{equation}
fluid stress-energy
\begin{equation}
\left(T_{\alpha}^{\beta}\right)_{;\,\beta} = G_{\alpha},
\label{fstress}
\end{equation}
and radiation stress-energy
\begin{equation}
\left(R_{\alpha}^{\beta}\right)_{;\,\beta} = -G_\alpha,
\label{rstress}
\end{equation}
together with the induction equation for the magnetic fields. In equations (\ref{fstress}) and (\ref{rstress}), $G_\alpha$ is the radiation 4-force density, which couples the fluid and radiation fields \citep{mihalas}. This term includes normal scattering, absorption, and emission, as well as thermal Comptonization of the radiation. We do not include the relativistic corrections to thermal Comptonization, as we make the simplifying assumption that Compton scattering is symmetric in the fluid frame (hence, there is no associated momentum exchange).  The form of $G^\alpha$ is
\begin{equation}
G^\alpha = A_1R^{\alpha \nu}u_{\nu} + \left(A_2 R^{\mu \nu} u_\mu u_\nu + \kappa_\mathrm{P}^\mathrm{a} \rho a_R T_\mathrm{gas}^4\right) u^{\alpha}~,
\label{G}
\end{equation}
where
\begin{equation}
A_1 = -\rho \left(\kappa_\mathrm{R}^\mathrm{a} + \kappa_\mathrm{s}\right)~,
\end{equation}
\begin{equation}
A_2 = - \rho\left[ \kappa_\mathrm{s} + 4 \kappa_\mathrm{s}\left(\frac{T_\mathrm{gas} - T_\mathrm{rad}}{m_e}\right) + \kappa_\mathrm{R}^\mathrm{a} - \kappa_\mathrm{J}^\mathrm{a} \right]~,
\end{equation}
$\kappa_\mathrm{s} = 0.34\,\mathrm{cm^2\,g^{-1}}$ is the opacity due to electron scattering, $\kappa^\mathrm{a}_\mathrm{R} = 1.6\times 10^{21} T^{-7/2}\rho\,\mathrm{cm^2\,g^{-1}}$ is the Rosseland mean of absorption, $\kappa_\mathrm{J}^\mathrm{a}$ is the J-mean of absorption, and $\kappa_\mathrm{P}^\mathrm{a}$ is the Planck mean of absorption.  We assume a nearly Planck spectrum, such that $\kappa_\mathrm{J}^\mathrm{a} = \kappa_\mathrm{P}^\mathrm{a} = 6.4\times 10^{22}T^{-7/2}\rho\,\mathrm{cm^2\,g^{-1}}$. Here, $T_\mathrm{rad}$ and $T_\mathrm{gas}$ are the radiation and gas temperatures, respectively. Note that we do not solve independently for the temperature of the electrons, but simply assume it is equal to the temperature of the plasma.  This should be sufficient in strongly coupled systems like the ones we simulate here, although in lower luminosity systems this does not hold \citep{ressler2015}.

The form of the conservation equations that we actually solve can be written as 
\begin{equation}
\partial_t D + \partial_i\left(DV^i\right) = 0~,
\end{equation}
\begin{equation}
\partial_t\mathcal{E} + \partial_i\left(-\sqrt{-g}T^i_t\right) = -\sqrt{-g}T^\alpha_\beta \Gamma^\beta_{t\alpha} - \sqrt{-g}G_t~,
\end{equation}
\begin{equation}
\partial_t\mathcal{S}_j + \partial_i\left(\sqrt{-g}T^i_j\right) = \sqrt{-g}T^\alpha_\beta \Gamma^\beta_{j\alpha} + \sqrt{-g}G_j~,
\end{equation}
\begin{equation}
\partial_t\mathcal{R} + \partial_i\left(\sqrt{-g}R^i_t\right) = \sqrt{-g}R^\alpha_\beta \Gamma^\beta_{t\alpha} - \sqrt{-g}G_t~,
\end{equation}
\begin{equation}
\partial_t\mathcal{R}_j + \partial_i\left(\sqrt{-g}R^i_j\right) = \sqrt{-g}R^\alpha_\beta \Gamma^\beta_{j\alpha} - \sqrt{-g}G_j~,
\end{equation}
and
\begin{equation}
\partial_t\mathcal{B}^j  + \partial_i\left(\mathcal{B}^j V^i - \mathcal{B}^i V^j\right) = 0~,
\end{equation}
where $D = W\rho$ is the generalized fluid density, $W = \sqrt{-g}u^t$ is the generalized boost, $V^i = u^i/u^t$ is the fluid transport velocity, $\mathcal{E} = -\sqrt{-g}T^t_t$ is the total energy density, $\mathcal{S} = \sqrt{-g}T^t_j$ is the covariant momentum density, $\mathcal{R} = \sqrt{-g}R^t_t$ and $\mathcal{R}_j = \sqrt{-g}R^t_j$ are the conserved radiation energy density and momentum, respectively, and $\mathcal{B}^j = \sqrt{-g}B^j$ is the boosted magnetic field three-vector. The magnetic field, $B^i = {}^*F^{\alpha i}$, is related to the co-moving field by
\begin{equation}
B^i = u^0b^i - u^i b^0~,
\end{equation}
where ${}^*F^{\alpha\beta}$ is the dual of the Faraday tensor.  These equations are solved using the explicit-implicit scheme described in \citet{fragile2014}. 

\section{Results}

In this section, we present results of our two main simulations. In the first, we find that our geometrically-thin, optically-thick, radiation-pressure-dominated disc collapses vertically on roughly the cooling timescale. The second simulation of an apparently stable, gas-pressure-dominated disc of similar height supports our claim that the first result is not a spurious numerical result.  

\subsection{Diagnostics}
Each simulation is post-processed in order to extract the thermodynamic and geometric properties of the disc. We mainly use density-weighted shell- and time-averaged quantities, as well as spacetime diagrams to present our results. A general expression for the density-weighted shell-average of a quantity is given by the following expression:
\begin{equation}
\langle f\rangle_\rho = \frac{\int\int f \sqrt{-g}\rho(R,\phi,z)dA_R}{\int\int \sqrt{-g}\rho(R,\phi,z)\,dA_R}~,
\label{weighted}
\end{equation}
where $dA_R$ is an area element normal to the radial direction. A time average of this quantity is computed as
\begin{equation}
\langle f\rangle_{\rho t} = \frac{\int\int\int f \sqrt{-g}\rho(R,\phi,z,t)dA_R dt}{\int\int\int \sqrt{-g}\rho(R,\phi,z,t)\,dA_R dt}~.
\label{tweighted}
\end{equation}
For the height of the disc we use a density-squared weighting:
\begin{equation}
\langle H\rangle_\rho = \sqrt{\frac{\int^{\pi/2}_0\int^{z_{max}}_{z_{min}} \sqrt{-g}\rho^2 (z - z_0)^2 dA_R}{\int^{\pi/2}_0\int^{z_{max}}_{z_{min}} \sqrt{-g}\rho^2 dA_R}}~,
\label{heightexpres}
\end{equation}
as it agrees better with our target height than other possible expressions, where $z_0 = 0$ represents the disc mid-plane.

\begin{figure*}
\centering
\includegraphics[width=2\columnwidth]{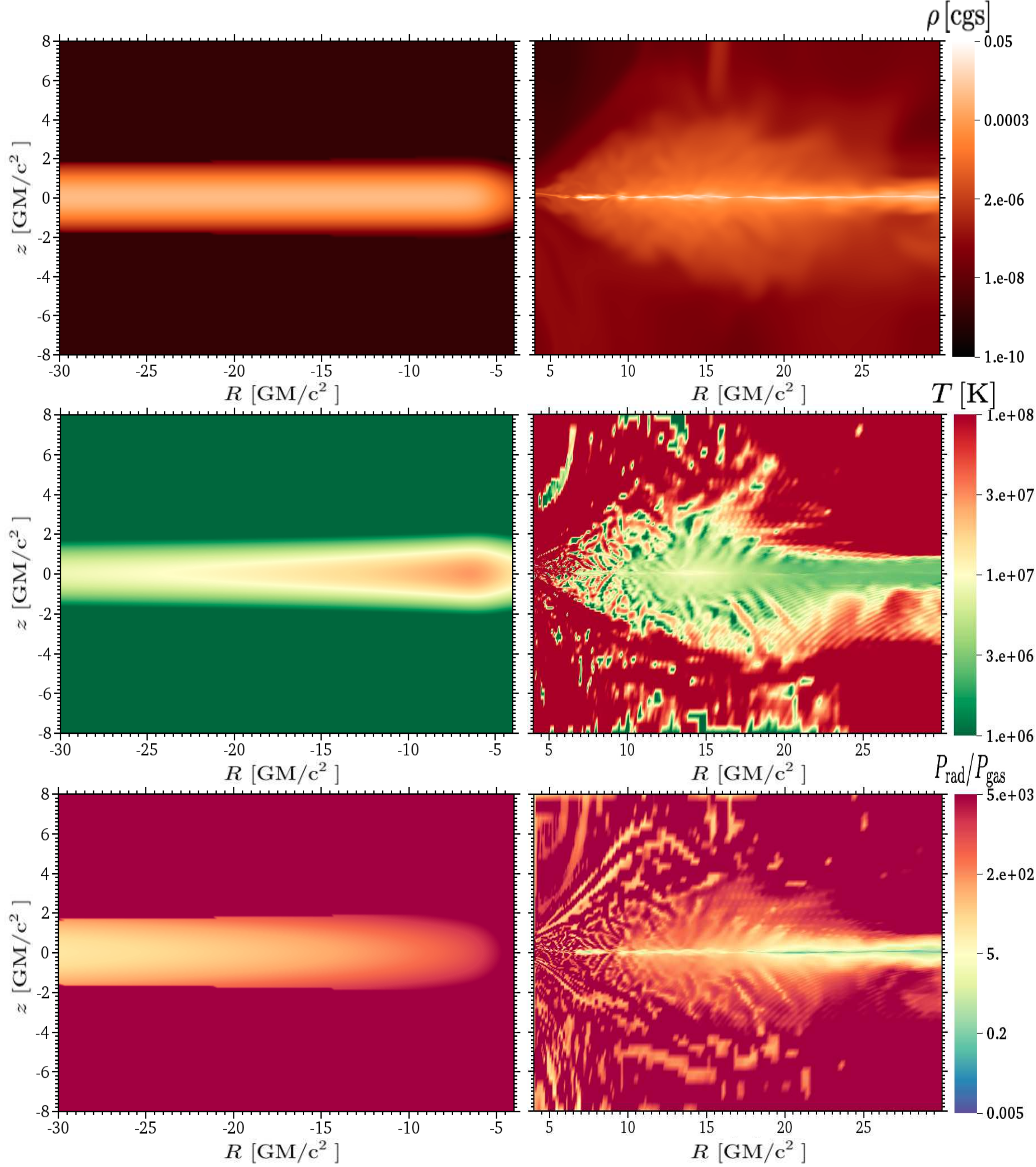}
\caption{An $R-z$ slice of three-dimensional simulation, RADPHR. The left and right panels correspond, respectively, to the initial and final stages of the simulation. From top to bottom, the panels show mass density, gas temperature, and the ratio of radiation pressure to gas pressure.}
\label{initial_final}
\end{figure*}

\begin{figure*}
\centering
\includegraphics[width=2\columnwidth]{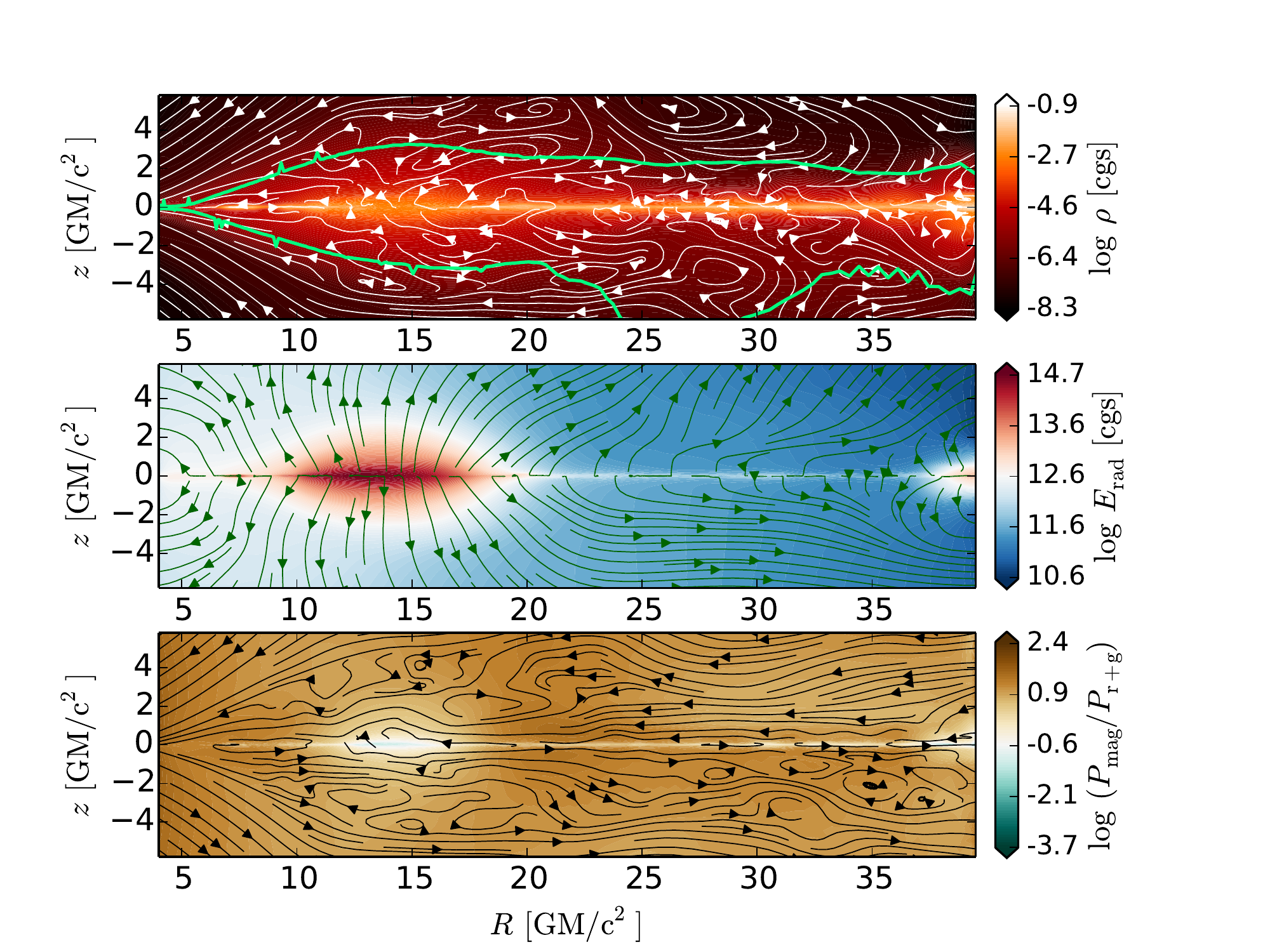}
\caption{An $R-z$ slice of the three-dimensional simulation, RADPHR. The panels show time averages over the last 10 ISCO orbital periods of the simulation. The top panel shows mass density, with streamlines of azimuthally averaged fluid velocity vectors. The photosphere is shown as a solid green curve. The middle panel shows the radiation energy, with streamlines of azimuthally averaged radiation velocity vectors. The lowest panel shows the ratio of magnetic pressure to the sum of radiation and gas pressures, with lines of azimuthally averaged magnetic field.}
\label{radphr_collapse}
\end{figure*}

We also track the photosphere of each disc, which we define as the $\tau = 1$ surface.  This is obtained by integrating the quantity $\kappa\rho$ from the lowest value of the $z$ coordinate  on the grid, $z_{\mathrm{min}}$, to the height where $\tau = 1$ and similarly from the highest value, $z_{\mathrm{max}}$, to the height where $\tau = 1$: 
\begin{equation}
\tau_< (z)= \int^z_{z_\mathrm{min}}u^t\kappa\rho\sqrt{g_{zz}}dz, \hspace{0.2in} \tau_> (z)=\int^{z_\mathrm{max}}_z u^t\kappa\rho\sqrt{g_{zz}}dz~,
\label{eq:tau}
\end{equation}
where $\kappa = \kappa_\mathrm{s} + \kappa^\mathrm{a}_\mathrm{R}$. We emphasize here that the photosphere is always well within our simulation domain.

The local heating rate per unit surface area owing to turbulence caused by the MRI is computed within the volume enclosed by the photosphere as
\begin{equation}
Q^+(R) = \frac{3}{2}\int\langle V^\phi W_{\hat{r}\hat{\phi}}\rangle_\phi dz ~,
\label{heatingrate}
\end{equation}
where the integration is carried out within the photosphere and the integrand is azimuthally averaged, with $V^\phi=\Omega$  the azimuthal component of the three velocity and $W_{\hat{r}\hat{\phi}}$  the contravariant $r$-$\phi$ component of the MHD stress tensor in the co-moving frame.  The radiative cooling is computed by tracking the radiative flux through the photosphere for each cylindrical shell:
\begin{equation}
Q^-(R) = \langle F^z_\mathrm{photo+}(R)\rangle_\phi - \langle F^z_\mathrm{photo-}(R)\rangle_\phi~,
\label{coolingrate}
\end{equation}
where $F^z_{\mathrm{photo}}(R) = -4/3 E_R u_R^z (u_R)_t$ is the flux escaping through the top or bottom photosphere. As advective cooling is not important in these simulations, we ignore its contribution to $Q^-$.  Also neglected is the contribution of the radial component of the radiative flux to cooling, which is appreciable only close to the ISCO.
\begin{figure}
\centering
\includegraphics[width=1\columnwidth]{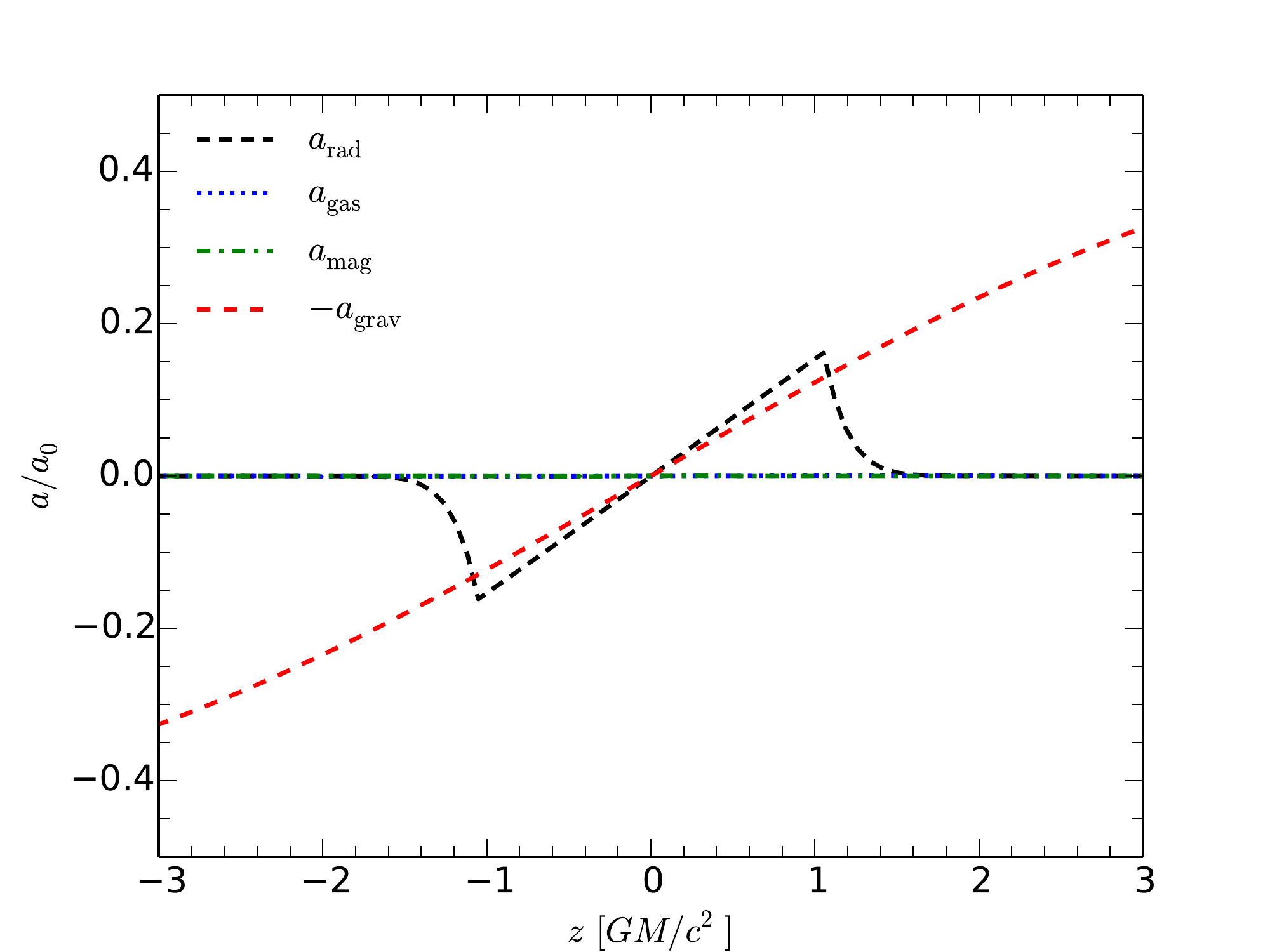}
\includegraphics[width=1\columnwidth]{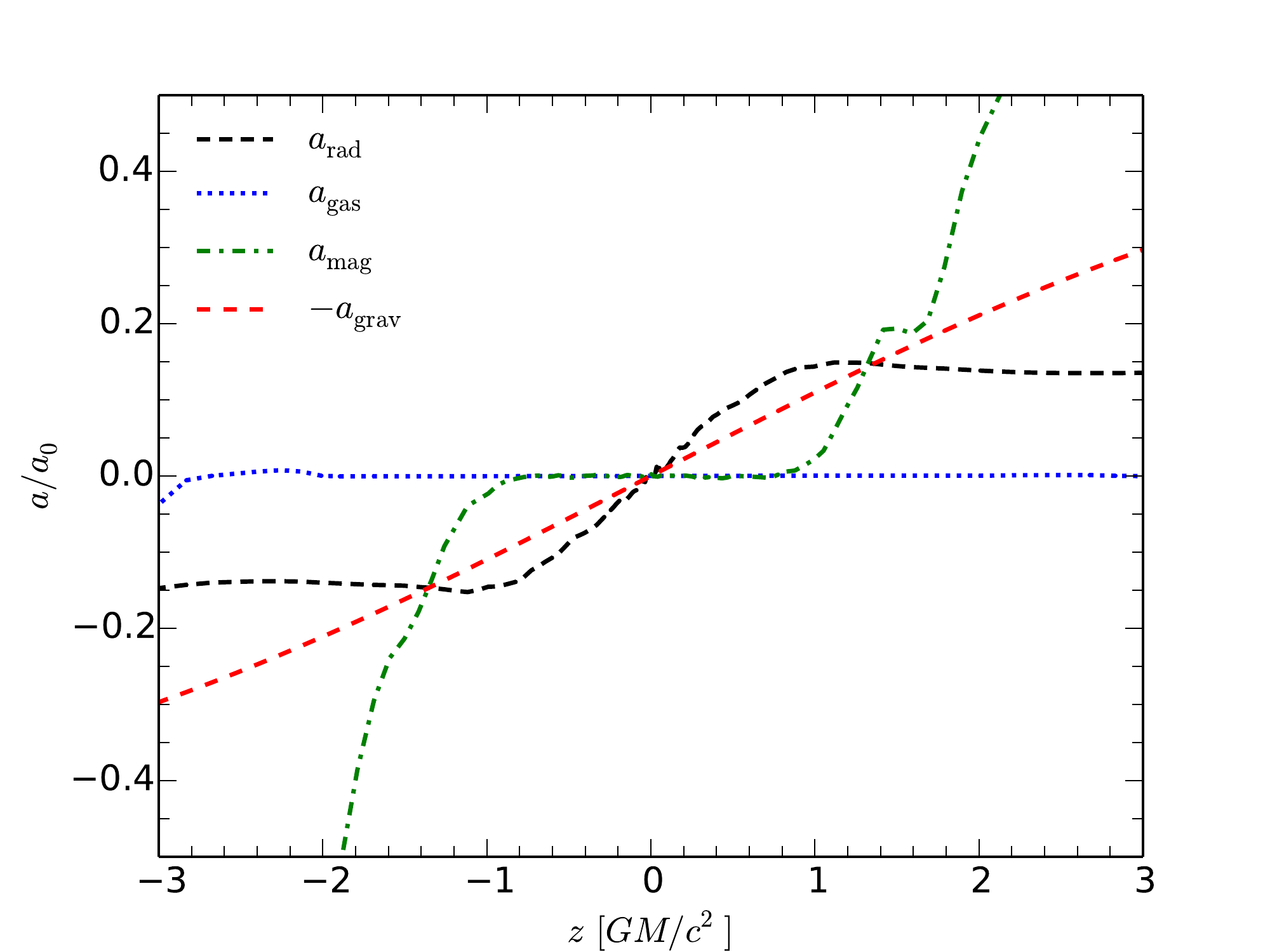}
\caption{Vertical profiles of the azimuthally averaged vertical acceleration components of radiation ($a_\mathrm{rad}$), gas pressure ($a_\mathrm{gas}$), and magnetic pressure ($a_\mathrm{mag}$) at $R = 10\,GM/c^2$, compared with the negative of the acceleration of gravity ($-a_\mathrm{grav}$).  The top panel shows the initial setup of the RADPHR simulation, while the data in the bottom panel are time averaged over the period from 1500 to $2500\,GM/c^3$.  All curves are normalized by  $a_{0}\equiv GM/r^2$.}
\label{accel}
\end{figure}
\subsection{Radiation-pressure-dominated disc} 

The goal of this setup (RADPHR/RADPLR) is to study the thermal stability of a radiation-pressure-dominated disc. Before addressing the thermal stability, though, we present a general overview of this simulation. 

 Fig.~\ref{initial_final} shows $R$-$z$ slices of the initial (left) vs. final (right) distribution of mass density (upper panel), temperature (middle panel), and the ratio of radiation pressure to gas pressure (lowest panel) for the RADPHR simulation. We clearly see in the upper-right panel that the disc collapses to a very thin structure. The middle-right panel shows that the disc cools down by about half an order of magnitude with respect to its initial gas temperature.  This panel, especially, shows small, wave-like structures outside the main body of the disc.  Those features are symptoms of primitive solver failures in the low-density background gas.  In these cases, the internal energy that is used to calculate the temperature and gas pressure can introduce such artificial features.  These errors do not appear to strongly influence the evolution of the disc itself.  The lower-right panel shows that during the entire simulation the radiation pressure remains dominant within the disc. 

In Fig.~\ref{radphr_collapse} we show additional $R$-$z$ slices of this simulation, now time averaged over the last 10 ISCO orbital periods ($t_\mathrm{ISCO} = 92.3\,GM/c^3$). The top panel shows mass density again with streamlines of the azimuthally averaged fluid velocity. The photosphere is also shown with green solid curves. Again we see that the disc has collapsed to a very thin height. 
In this case, we terminated the simulation where we did, because we can no longer resolve the disc. The middle panel shows the radiation energy  with streamlines of the azimuthally averaged radiation velocity (corresponding to the direction of the radiative flux vectors). We see in this panel that the radiation escapes mostly vertically from the disc, which is expected for geometrically thin discs.  The lowest panel shows the ratio of the magnetic pressure to the sum of the radiation and gas pressures, $P_\mathrm{mag}/(P_\mathrm{rad} + P_\mathrm{gas})$, with lines of the azimuthally averaged magnetic field. There is a definite radial structure to the magnetic field lines in the background gas, while the field is more turbulent near the disc mid-plane.  Magnetic pressure never dominates in the body of the disc during the entire simulation; thus, we do not expect the disc to be stabilized by the magnetic fields.  

Fig.~\ref{accel} shows the vertical acceleration components
\begin{eqnarray}
a_\mathrm{grav} & = &
-zM(1-2M/R)^{-1}/(R^2 + z^2)^{3/2} ~, \nonumber \\
a_\mathrm{rad} & = & G_z/\rho ~, \nonumber \\
a_\mathrm{gas} & = & -(\nabla P_\mathrm{gas})_z/\rho ~\mathrm{,~and} \nonumber \\
a_\mathrm{mag} & = & -(\nabla P_\mathrm{mag})_z/\rho 
\end{eqnarray}
in the coordinate frame. We can see that at the beginning of the simulation, as well as during the subsequent collapse, the disc is approximately in hydrostatic equilibrium. This demonstrates that the collapse is not caused by a loss of hydrostatic equilibrium.  It is only after the collapse that hydrostatic equilibrium is violated (for $\vert z \vert < 1$). We also note in the bottom panel that magnetic pressure dominates and even exceeds gravity above and below the disc. As we will see, this additional pressure support allows the photosphere to be outside the main body of the disc.

\begin{figure}
\centering
\includegraphics[width=1\columnwidth]{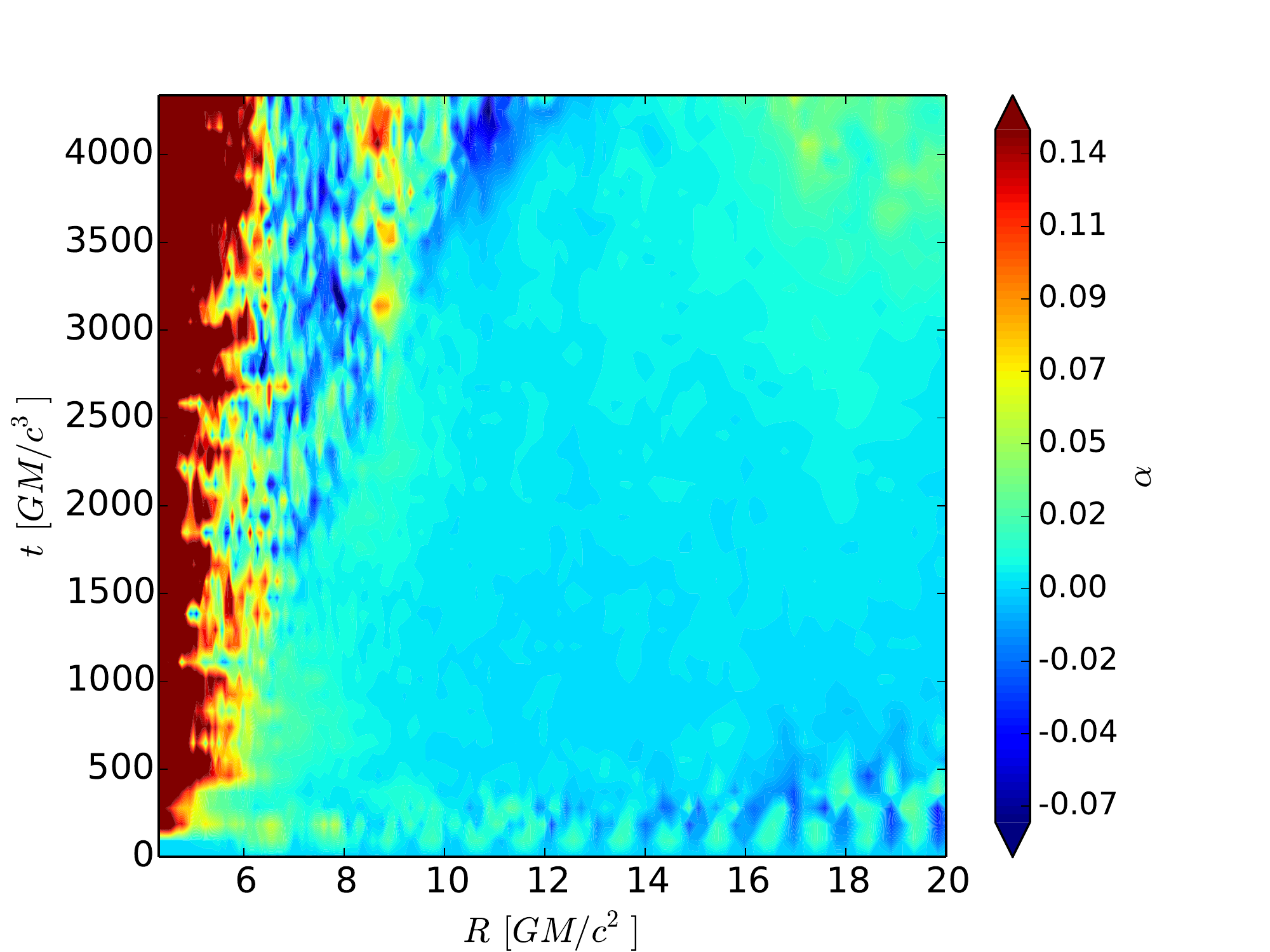}
\caption{Space time plot of density-weighted, shell-averaged viscosity parameter, $\alpha \equiv \langle W_{\hat{r}\hat{\phi}}/P_\mathrm{tot}\rangle_\rho$.}
\label{alpha_radphr}
\end{figure}

Fig.~\ref {alpha_radphr} shows the viscosity parameter, defined here as the density-weighted, height-averaged ratio of the (covariant) $\hat{r}$-$\hat{\phi}$ component of the stress tensor to the total pressure $\alpha\equiv \langle W_{\hat{r}\hat{\phi}}/P_\mathrm{tot}\rangle_\rho$. We find a nearly constant and uniform value of $\alpha=0.02.$

\begin{figure}
\centering
\includegraphics[width=1\columnwidth]{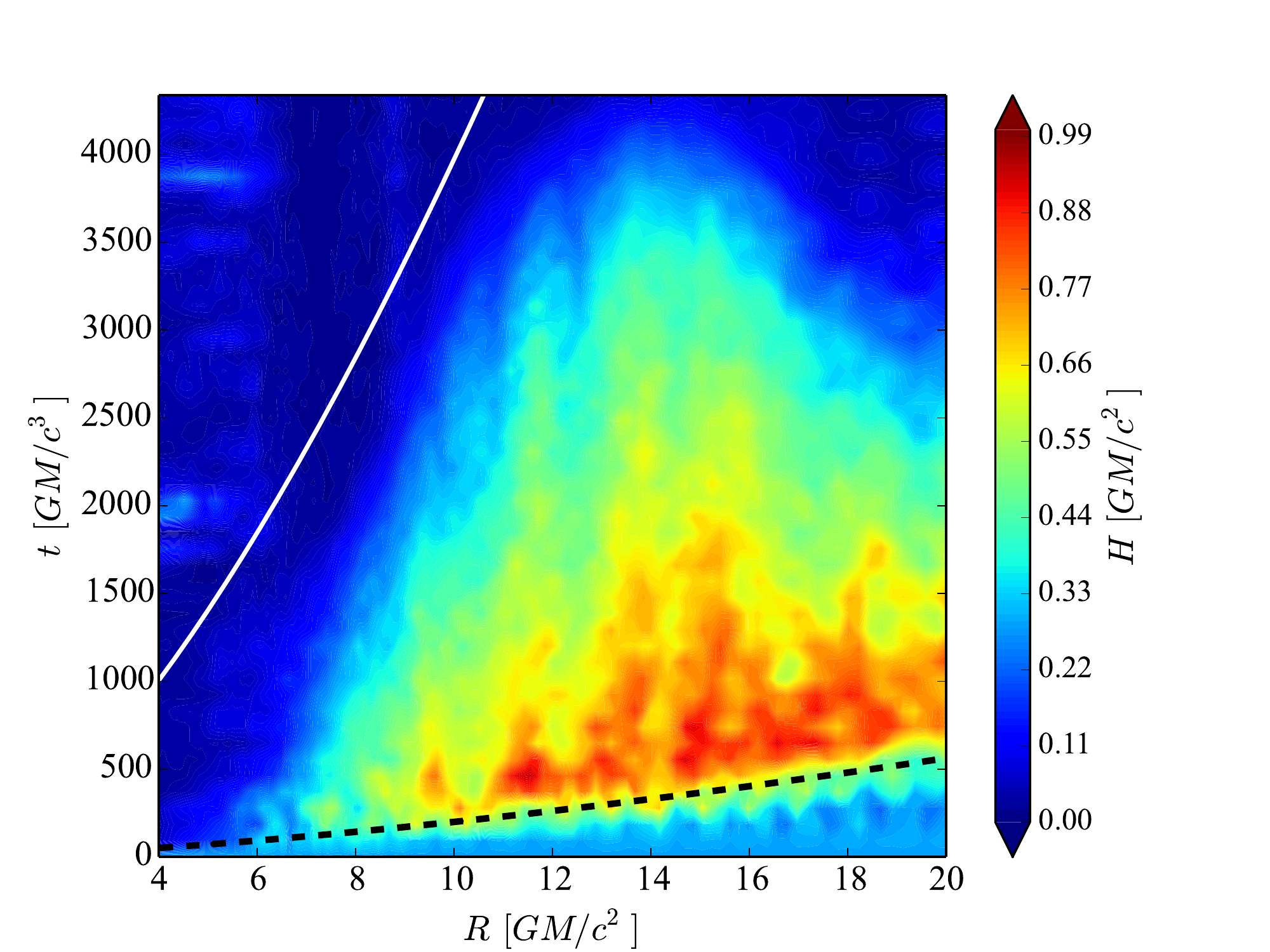}
\caption{Spacetime plot for the azimuthally averaged, density-weighted height (\ref{heightexpres}) of the disc in the high-resolution, RADPHR simulation. The dashed curve shows the local MRI growth time, while the solid curve shows the estimated cooling time of an equilibrium disc, $(\alpha\Omega)^{-1}$.}
\label{heightRS}
\end{figure}

\begin{figure}
\centering
\includegraphics[width=1\columnwidth]{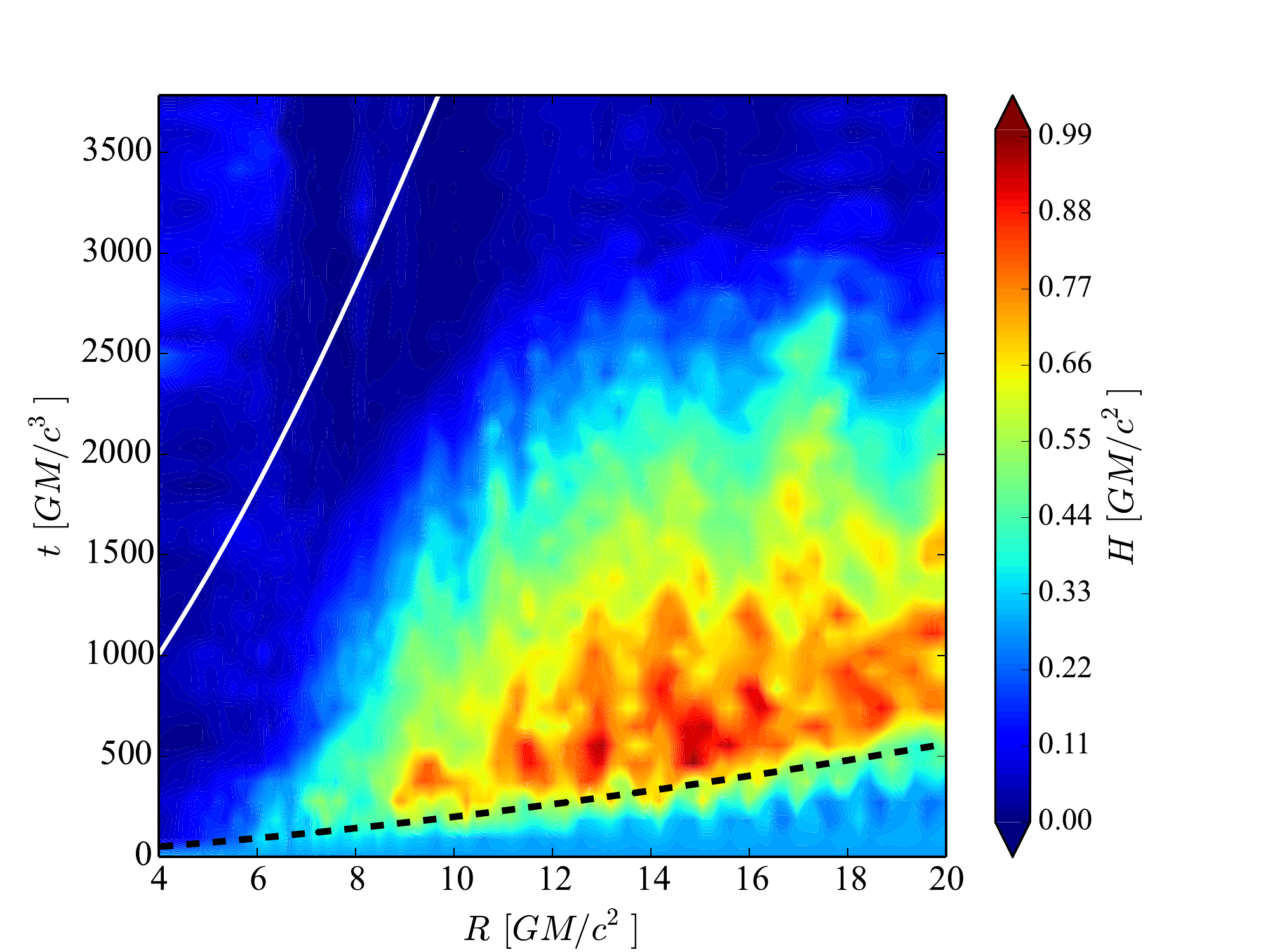}
\caption{Same as Fig. \ref{heightRS}, but for the low-resolution, RADPLR case.}
\label{heightRSLR}
\end{figure}
\begin{figure}
\centering
\includegraphics[width=1\columnwidth]{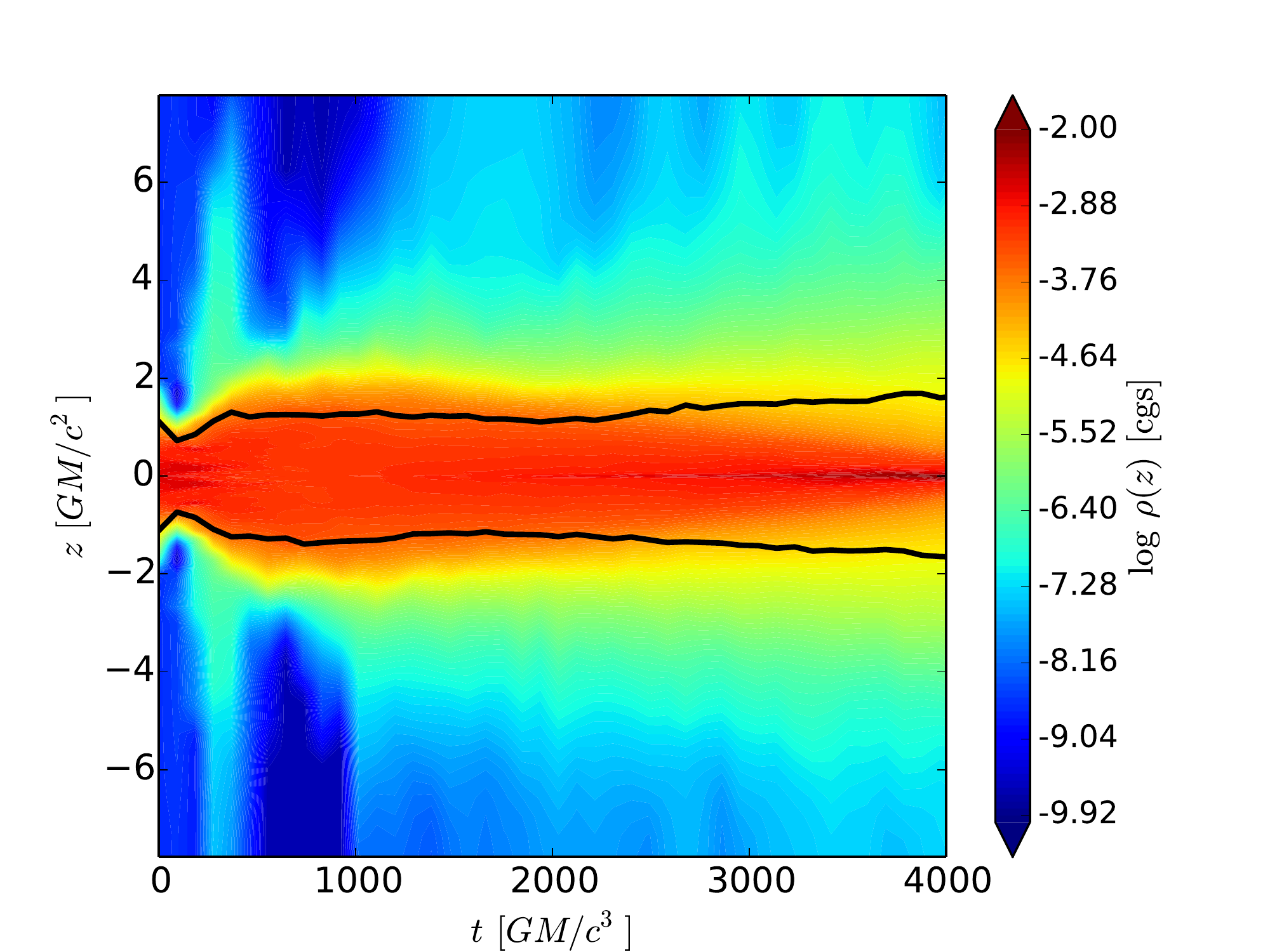}
\caption{Spacetime plot of the vertical profile of the radially $(R < 20 r_g)$ and azimuthally averaged mass density for the collapsing disc in the RADPHR simulation. The black curves show the radially and azimuthally averaged photospheric height.}
\label{vertmassD}
\end{figure}

In Fig.~\ref{heightRS} we show a spacetime plot of the azimuthally averaged radial profile of disc height, computed using equation (\ref{heightexpres}). The figure shows that the height initially increases after one local MRI growth time. Subsequently, the disc collapses on a timescale comparable to the local equilibrium cooling time (solid white curve), which is estimated from thin disc theory to be $t_\mathrm{cool}(R) = 2\pi/(\alpha \Omega)$. The disc collapse negates the initial expansion such that the final height is at least a factor of five smaller than the initial one. We show a similar spacetime plot for the height of our low-resolution, RADPLR case in Fig.~\ref{heightRSLR}. Comparing Figs. \ref{heightRS} and \ref{heightRSLR}, we see that it takes somewhat longer for the disc to collapse in our high-resolution simulation.  It could be that at even higher resolution we might find that the disc will take even longer to collapse.  This point will have to await future simulations.

The vertical collapse of the disc can also be seen in Fig. \ref{vertmassD}, which shows a spacetime plot of the radially and azimuthally averaged density as a function of height, $z$.  The narrowing of the high density region of the plot with time reflects the collapse of the disc.  As the disc is collapsing, the photosphere (black curve) remains at a relatively constant height. This is because there is a magnetically supported atmosphere at $|z|\ge 1 GM/c^2$ (see Fig. \ref{accel}). Recall, too, that the photosphere is calculated by integrating the opacity from the domain boundaries toward the mid-plane.

\begin{figure}
\centering
\includegraphics[width=1\columnwidth]{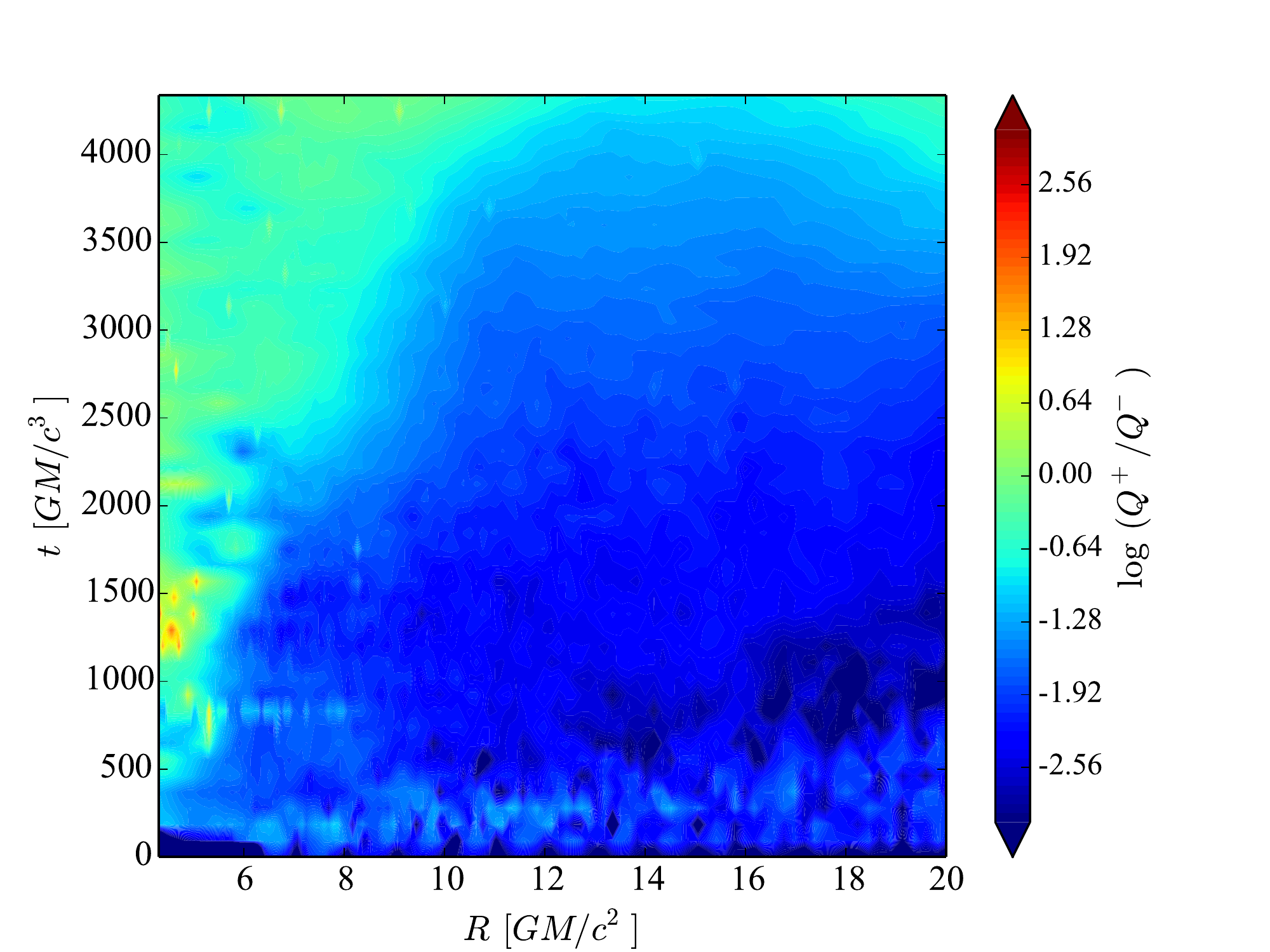}
\caption{Spacetime plot for the ratio of the heating rate, $Q^+$ (\ref{heatingrate}), to the cooling rate, $Q^-$ (\ref{coolingrate}), for the RADPHR simulation.}
\label{heatcoolRS}
\end{figure}

Another clue to the thermal state of the disc comes from looking at the time evolution of the heating and cooling rates. In Fig.~\ref{heatcoolRS}, we show a spacetime plot of the ratio of the heating rate to the cooling rate, computed using equations (\ref{heatingrate}) and (\ref{coolingrate}), respectively. The important takeaway is that cooling dominates over heating everywhere in the disc until after the disc has collapsed. This is even true at the outset of the simulation.  As we noted, this simulation does not begin in thermal equilibrium. Heating does appear to balance cooling inside the ISCO, but this has little impact on our results.

\begin{figure}
\centering
\includegraphics[width=1\columnwidth]{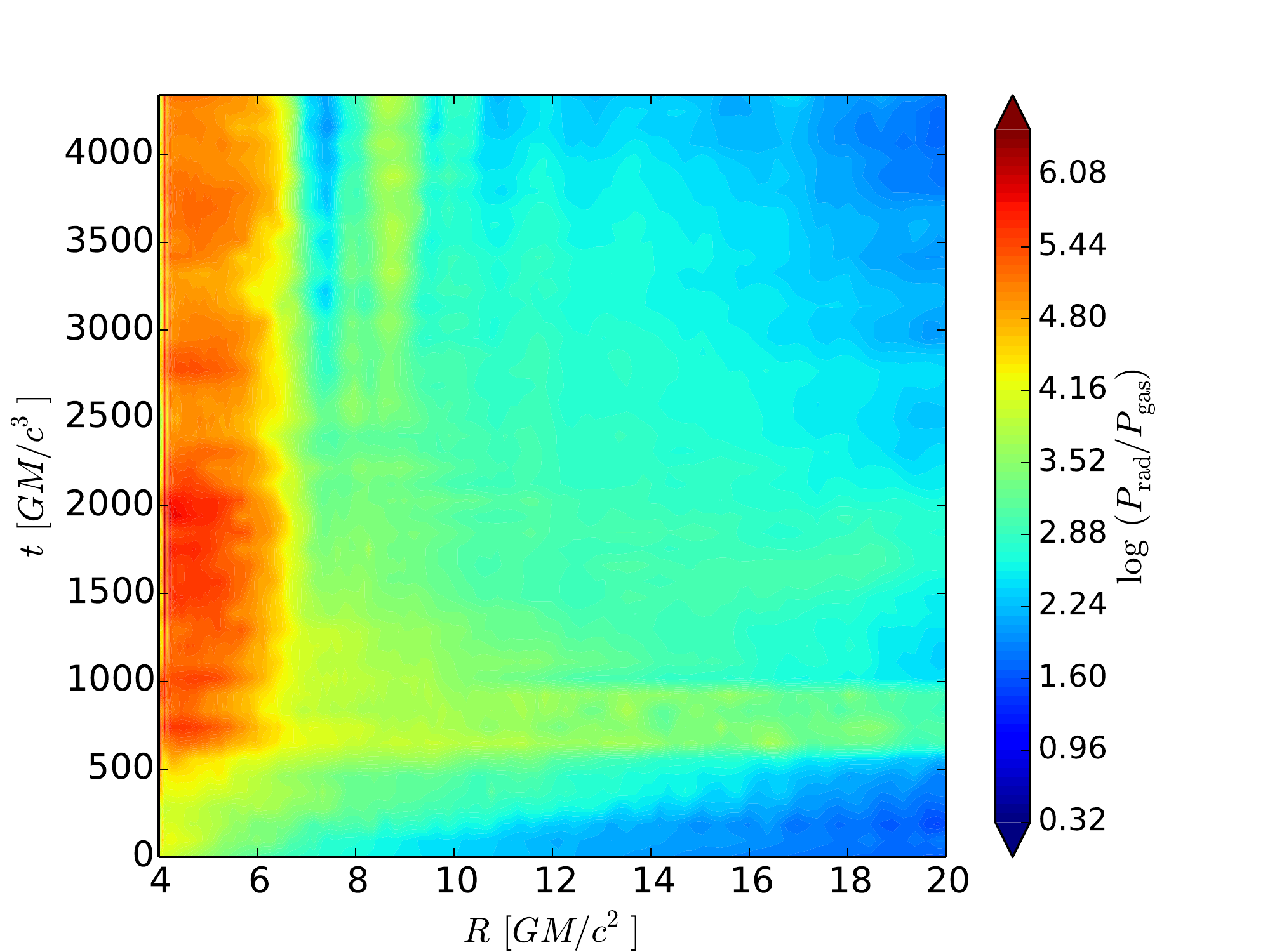}
\caption{Spacetime plot of the ratio of the density-weighted, shell-averaged radiation pressure, $P_{\mathrm{rad}}$, to gas pressure, $P_\mathrm{gas}$, for the RADPHR simulation.}
\label{pradpgas}
\end{figure}

As mentioned at the beginning of this section, the RADPHR and RADLR simulations are initialized with the disc being radiation pressure dominated. In Fig.~\ref{pradpgas} we show that it remains so throughout the simulation. Obviously the region inside the ISCO is strongly radiation pressure dominated.  This is not surprising as this region is largely devoid of gas, but filled with radiation (compare the top and middle panels of Fig. \ref{radphr_collapse}).

\begin{figure}
\centering
\includegraphics[width=1\columnwidth]{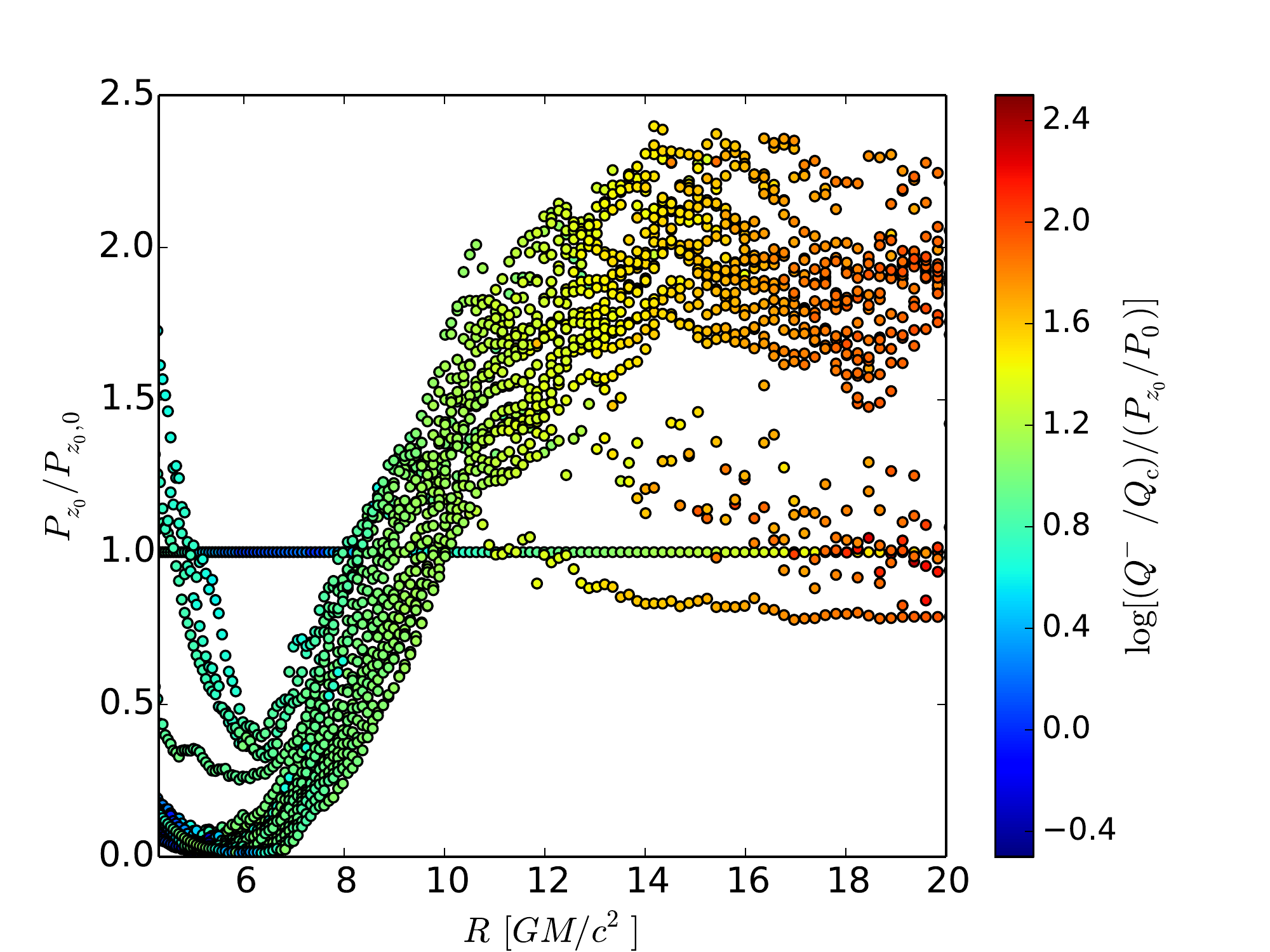}
\caption{Scatter plot of cooling rate, $Q^-$, normalized by $Q_\mathrm{c}(P_\mathrm{z_0}/P_0)$, where $Q_\mathrm{c} = c \Omega^2H_0/\kappa_\mathrm{s}$ with $H_0=0.4\,GM/c^2$, $P_\mathrm{z_0}$ is the total midplane pressure, and $P_0$ the initial total mid-plane pressure at $R=10\,GM/c^2$. Data for the first twenty orbits are shown.}
\label{coolpres}
\end{figure}

\begin{figure}
\centering
\includegraphics[width=1\columnwidth]{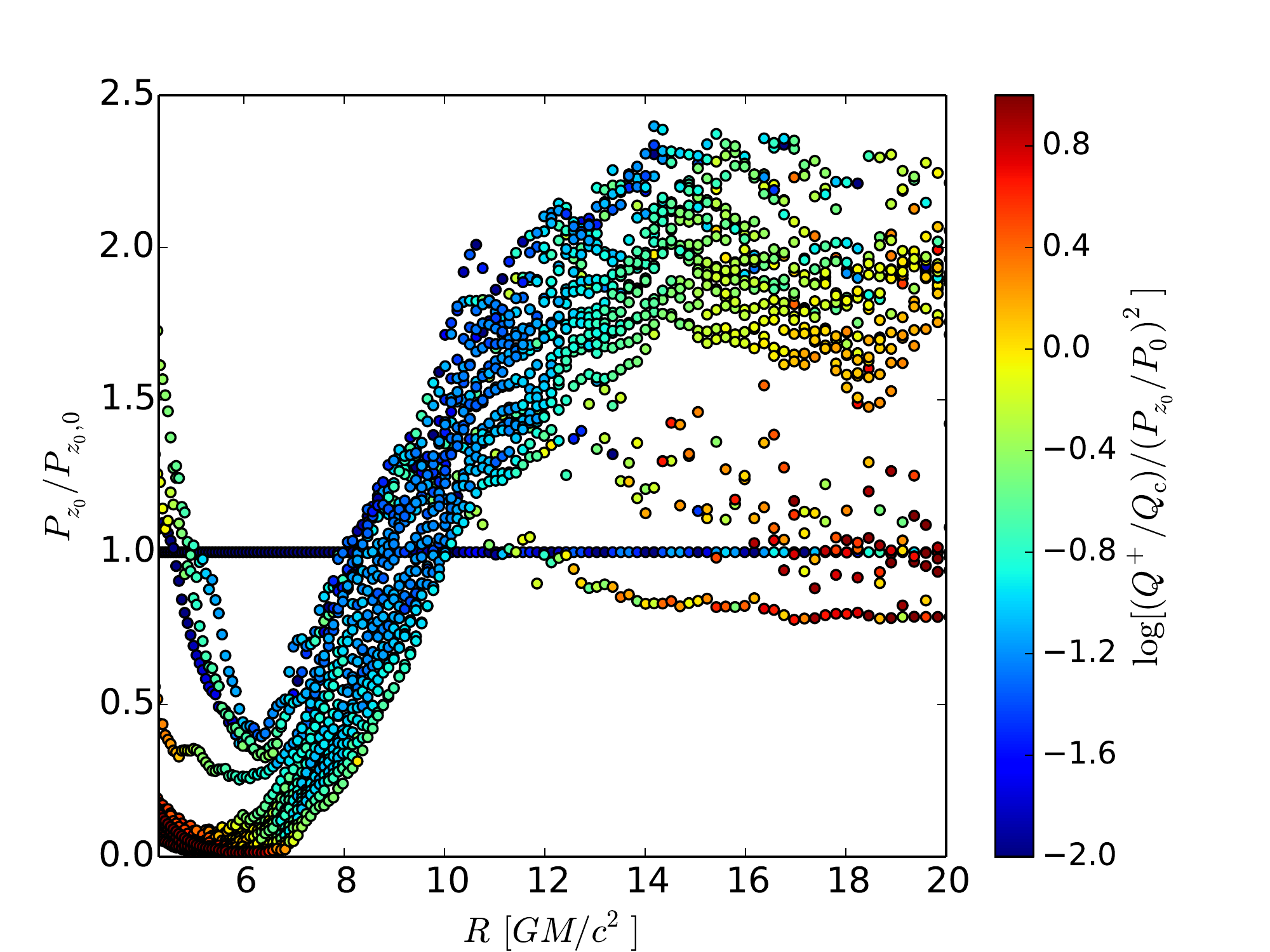}
\caption{Scatter plot similar to Fig. \ref{coolpres}, but for the heating rate, $Q^+$, normalized by $Q_\mathrm{c}(P_\mathrm{z_0}/P_0)^2$.}
\label{heatpres}
\end{figure}

To demonstrate a thermal instability, we need to show that the heating and cooling rates responding differently to changes in the total mid-plane pressure, which we do in Figs.~\ref{coolpres} and \ref{heatpres}. In particular, Fig.~\ref{coolpres}, shows that the cooling rate, $Q^-$, is proportional to the mid-plane pressure, as expected from the radiative diffusion equation. Note that $Q^-$ has been scaled in the figure with one power of total pressure.
At early times ($t < 2000\,GM/c^3$), and especially at small radii ($R < 12\,GM/c^2$), we see in Fig. \ref{heatpres} that the heating rate, $Q^+$, scales as the square of the mid-plane pressure. Note that in this figure,  $Q^+$ has been scaled by the square of total pressure. Thus, the dependence on pressure is steeper in the case of heating ($Q^+\propto P^2_{z_0}$) than in the case of cooling ($Q^-\propto P_{z_0}$), which is the hallmark of the thermal instability.  So, even if this simulation had started in thermal equilibrium, it would have still been prone to a thermal runaway.


At later times the discussion of heating and cooling, and relating the numerical results to classic theory of accretion discs, is complicated by the presence of two different scale-heights in the disc. The density scale height $H$ illustrated in Figs.~\ref{heightRS} and \ref{heightRSLR} is very different from the height of the photosphere in Fig.~\ref{vertmassD}. We remind the reader that the cooling rate $Q^-(R)$ is the flux leaving through the photosphere at radius $R$, and the heating rate $Q^+(R)$ is computed in a cylindrical shell of the same radius from the lower to the upper photosphere.

\begin{figure}
\centering
\includegraphics[width=1\columnwidth]{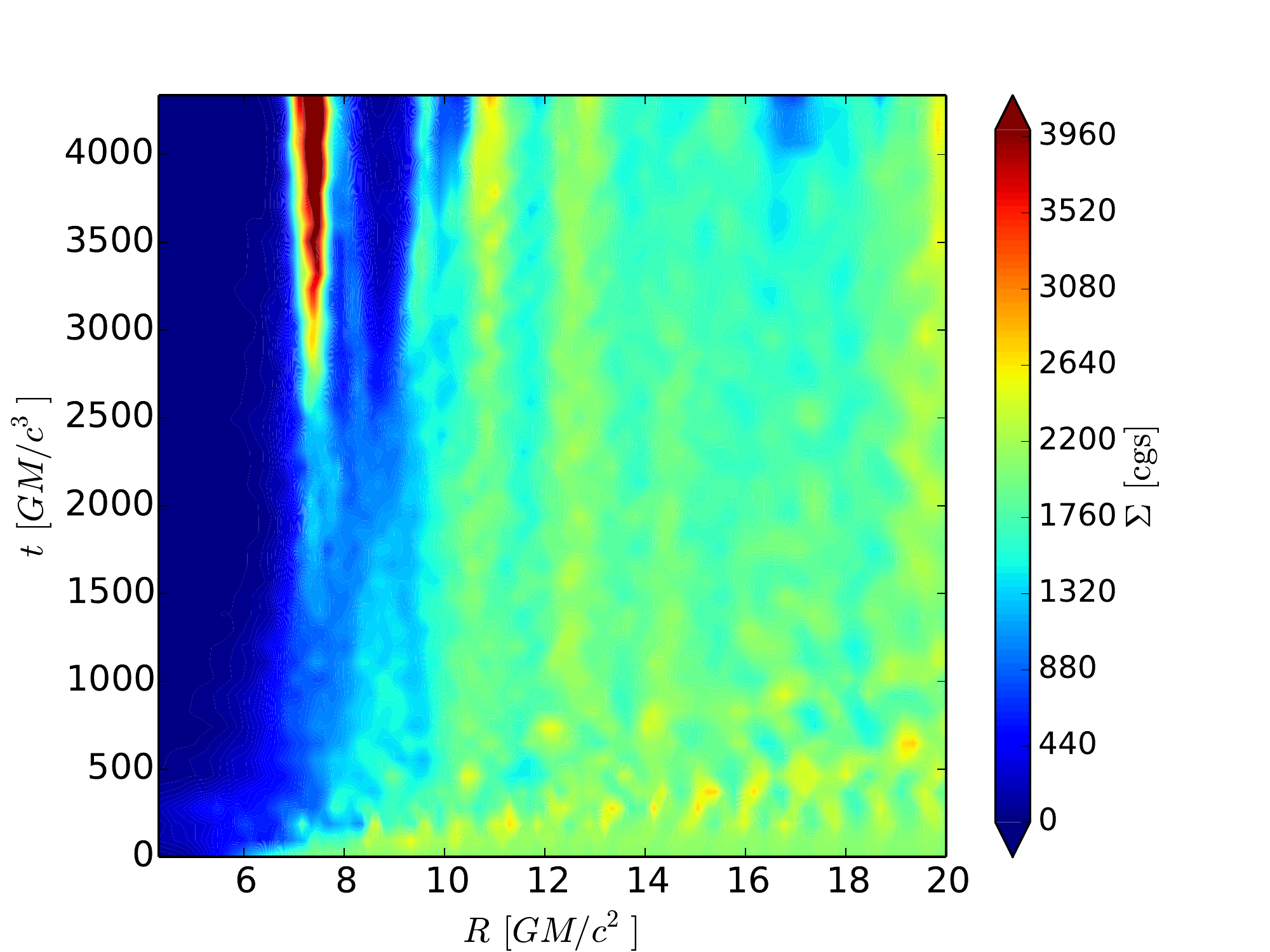}
\caption{Spacetime plot for the surface density, $\Sigma$, of the disc in the RADPHR simulation.}
\label{sigma}
\end{figure}

\begin{figure*}
\centering
\includegraphics[width=1.5\columnwidth]{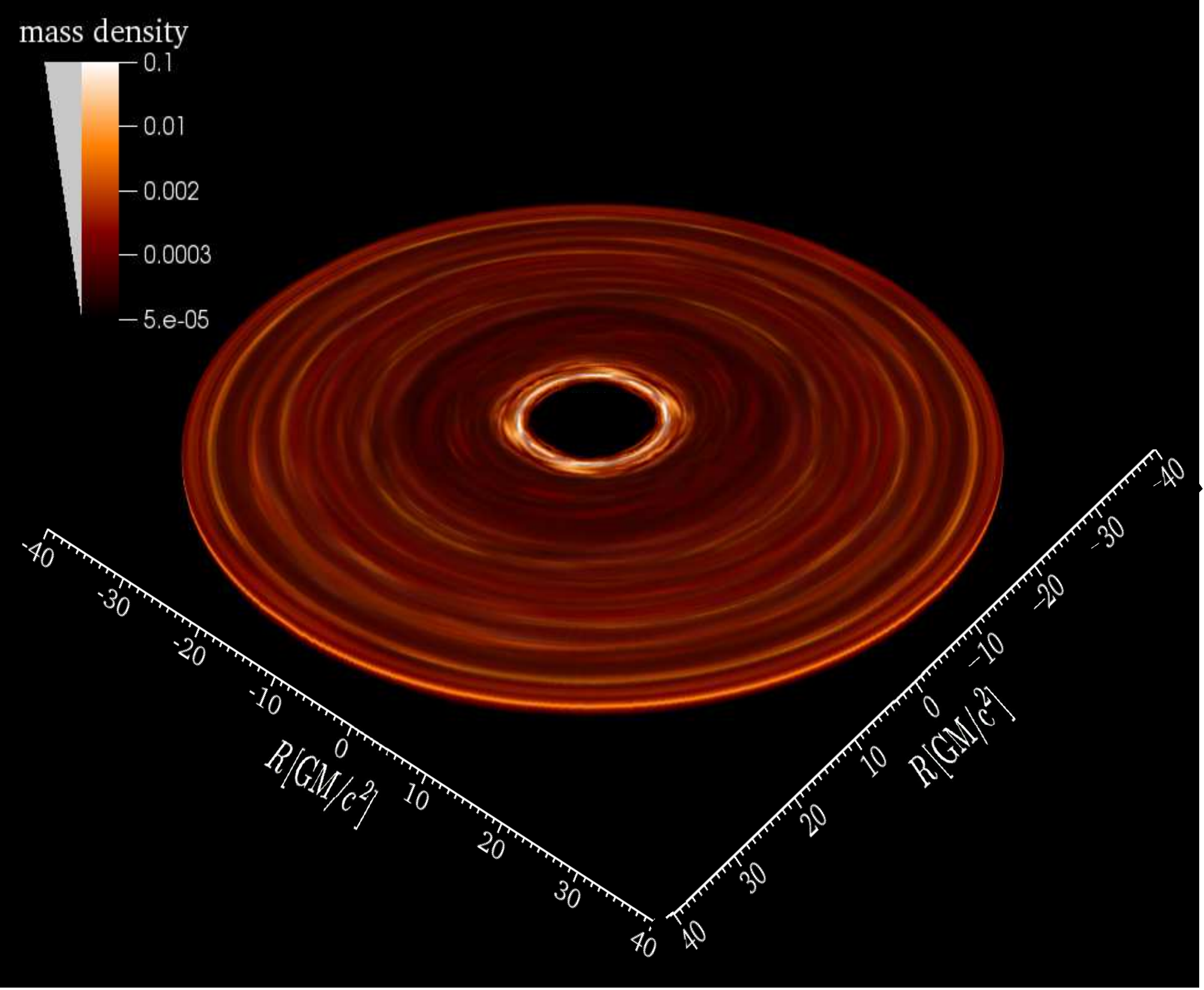}
\caption{Three dimensional volume visualization of disc mass density at $t = 3500\,GM/c^3$ of the RADPHR simulation. The actual simulation only covered $(0,{\rm \pi}/2)$ in azimuth, so has been reflected across two planes to show a full $2 {\rm \pi}$ version.}
\label{viscousi}
\end{figure*}

\begin{figure}
\centering
\includegraphics[width=1\columnwidth]{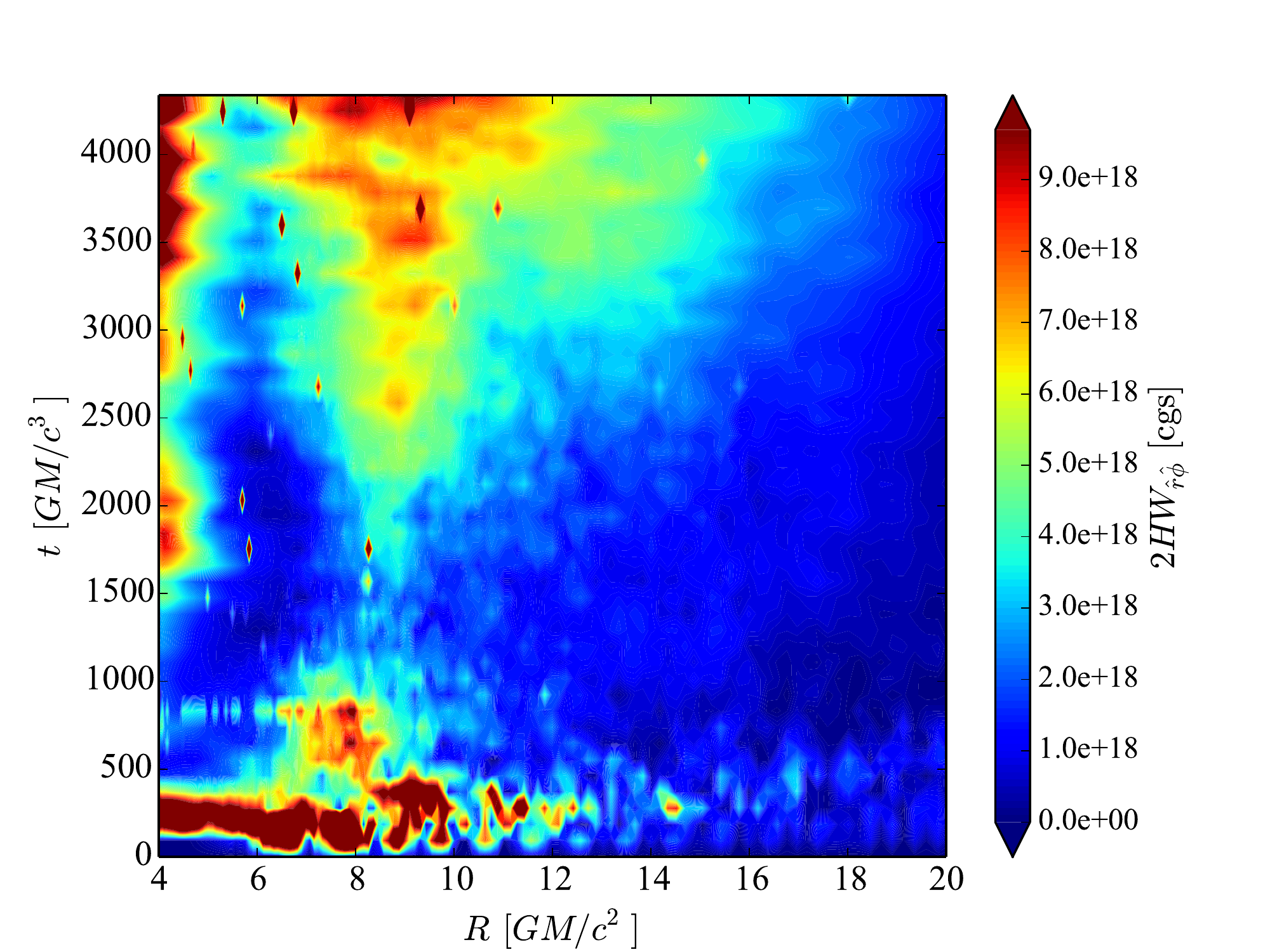}
\caption{Spacetime plot for the vertically integrated stress, $2HW_{\hat{r}\hat{\phi}}$, of the disc in the RADPHR simulation.}
\label{vertStress}
\end{figure}

In Fig.~\ref{sigma}, we show a spacetime plot of the surface density, $\Sigma$, of the disc. First, we note that $\Sigma$ remains nearly constant throughout the simulation. This is important as it demonstrates that the disc collapse is not caused by matter being drained into the black hole. However, a closer inspection of Fig. \ref{sigma} shows that the disc seems to collect into distinct rings after a time of $\sim 1000\,GM/c^3$.  For instance, a very dense ring forms at $R \approx 7\,GM/c^2$. As seen in Fig.~\ref{viscousi}, these really are rings and not spiral structures. This is very suggestive of the viscous instability \citep{lightman1974}. Additional rings are seen forming at $R \approx 11, 13$, and $15\,GM/c^2$.  Thus, the spacing between the rings in the inner disc is roughly the same as the height of the disc ($\Delta R\sim H$), as expected in the early stages of the viscous instability \citep{lightman1974}.  The anticipated growth timescale for the instability is roughly the viscous timescale multiplied by the relative size squared of the structures being formed, i.e. $t_\mathrm{LE} \sim (R/H)^2/(\alpha\Omega)\times(\Delta R/R)^2$.  For the innermost ring, this is approximately $900\,GM/c^3$, roughly consistent with the first appearance of the ring in Fig. \ref{sigma}.  If this collection of mass into rings is indeed owing to the viscous instability, it should also be accompanied by a rise in the viscous stress in the regions between the rings since $W_{\hat{r}\hat{\phi}} \propto \Sigma^{-1}$.  Comparing Figs. \ref{sigma} and \ref{vertStress}, we do see that the prominent low-$\Sigma$ gap between $R=8$ and $10\,GM/c^2$ in Fig.~\ref{sigma} corresponds to the highest stress region in Fig.~\ref{vertStress}, providing additional evidence that we may be seeing the viscous instability in a simulation for the first time.

\subsection{Gas-pressure-dominated disc}

In the previous section we described a radiation-pressure-dominated thin disc which is unstable to collapse. In this section, we contrast that result with a baseline, gas-pressure-dominated simulation. The primary goal in performing this simulation is to ensure that the result of the previous section is not a numerical artifact and that, indeed, our numerical setup is able to evolve stable thin disc configurations for sufficiently long duration. Here we focus on the low-resolution simulation, GASPLR, as it ran for significantly longer than its high-resolution counterpart. Even so, it did not run long enough to achieve inflow equilibrium through much of the radial domain. If the goal was really to understand the behavior of this disc, then this simulation should be run even longer.  However, we were able to run it for many thermal times, long enough to provide a convincing counter-example to our collapsing disc.

\begin{figure}
\centering
\includegraphics[width=1\columnwidth]{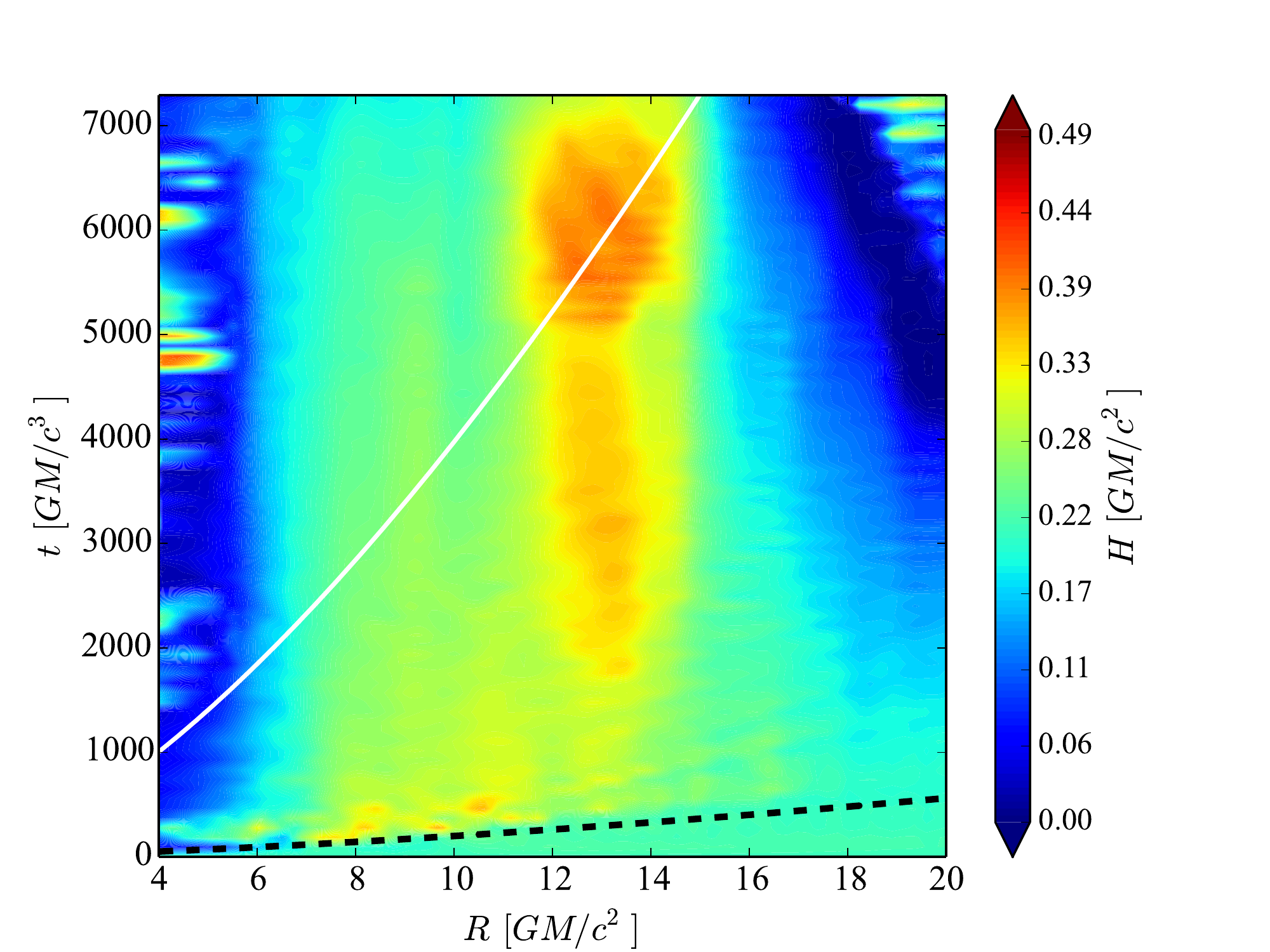}
\caption{Same as Fig. \ref{heightRS}, but for the GASPLR simulation.}
\label{hrvs}
\end{figure}

We start with a spacetime plot of the disc height (Fig.~\ref{hrvs}), comparable to Figs.~\ref{heightRS} and \ref{heightRSLR}. In it we see that the region inside of $R\approx 12\,GM/c^2$ maintains a nearly constant height following the onset of the MRI (dashed line). However, the GASPLR simulation does exhibit some thinning in its outer regions, an effect that was also seen in the radiation-pressure-dominated simulations. We suspect this is owing to the fact that the MRI is not sufficiently well resolved in those regions (see Sec. \ref{sec:caveats}), which allows cooling to dominate.

\begin{figure}
\centering
\includegraphics[width=1\columnwidth]{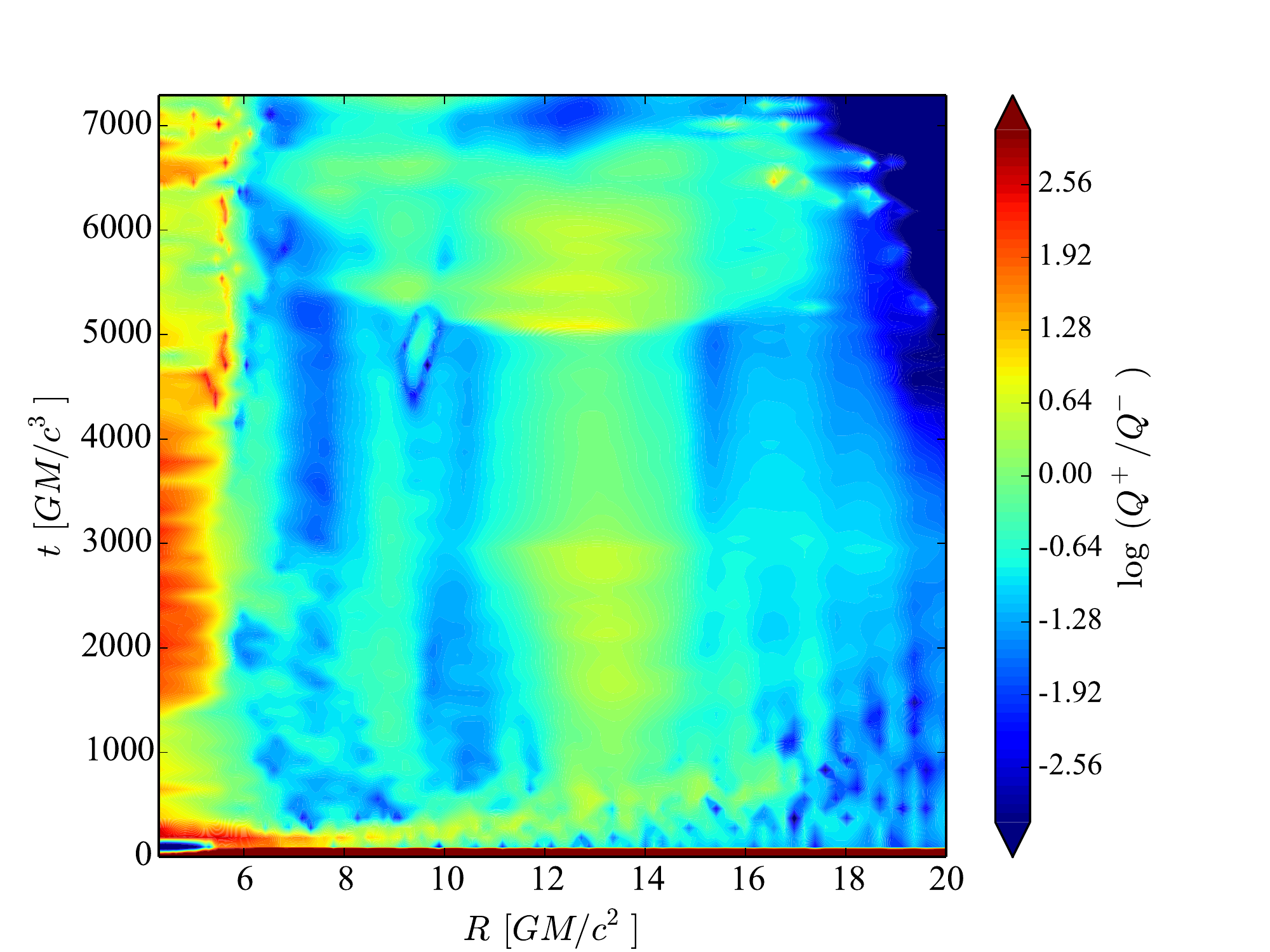}
\caption{Same as Fig. \ref{heatcoolRS}, but for the GASPLR simulation.}
\label{heat_cool_rvs}
\end{figure}

In Fig.~\ref{heat_cool_rvs} we show a space-time plot of the ratio of heating to cooling in the GASPLR case. We see that heating and cooling are nearly in balance over time over most of the disc. However, in the outer regions and at late times, there are parts of the disc where cooling dominates.  As with the collapsing scale height in these regions, we attribute this behavior to the under resolved MRI.

\begin{figure}
\centering
\includegraphics[width=1\columnwidth]{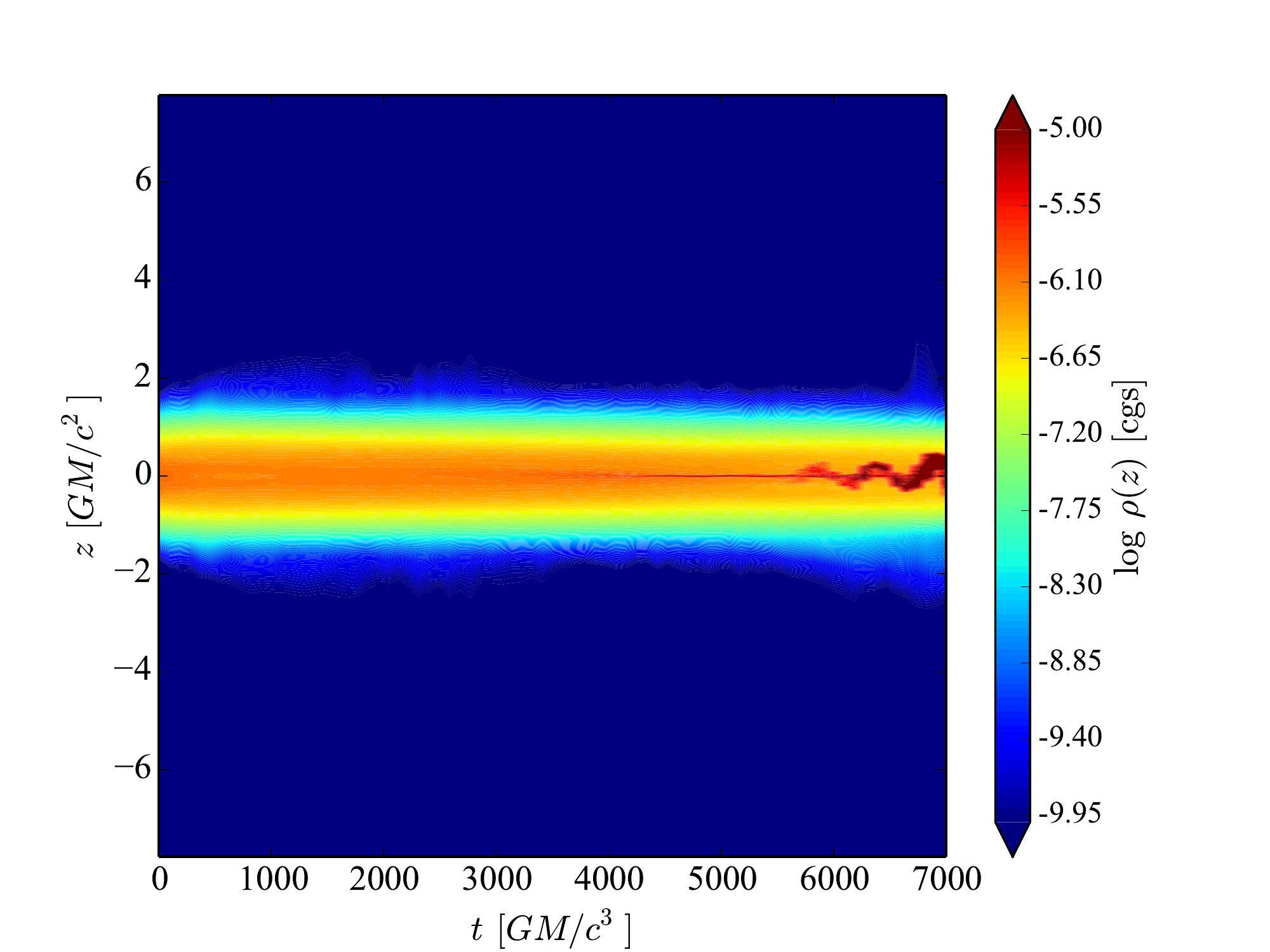}
\caption{Same as Fig. \ref{vertmassD}, but for the GASPLR simulation.  In this case, the disc is optically thin.}
\label{rvs_vert}
\end{figure}

In Fig.~\ref{rvs_vert} we show the vertical profile of the radially and azimuthally averaged mass density for the GASPLR simulation. As with Fig. \ref{hrvs}, we find that the gas-pressure-dominated disc maintains its height much better than the radiation-pressure-dominated one. The growth in the mid-plane density at late times is dominated by the thinning of the outer regions ($R > 16\,GM/c^2$) of the disc (also seen in Fig. \ref{hrvs}). 

\subsection{Caveats}
\label{sec:caveats}

One major concern in performing numerical simulations of very thin accretion discs is resolving the MRI. Multiple resolution studies have shown that nearly all global simulations done to date have been under-resolved in this regard \citep[e.g.][]{hawley2011,hawley2013}. A particular worry in our case is that under-resolving the MRI might cause the disc to collapse due to insufficient heating. Admittedly, as we show, some of our simulations fall below the ideal resolution. This is an unfortunate limitation of trying to perform global simulations, and doing so for thin discs only exacerbates the problem. However, an attempt to perform such thin disc simulation is necessary to make progress in understanding the physics of thin discs. 

\begin{figure}
\centering
\includegraphics[width=1\columnwidth]{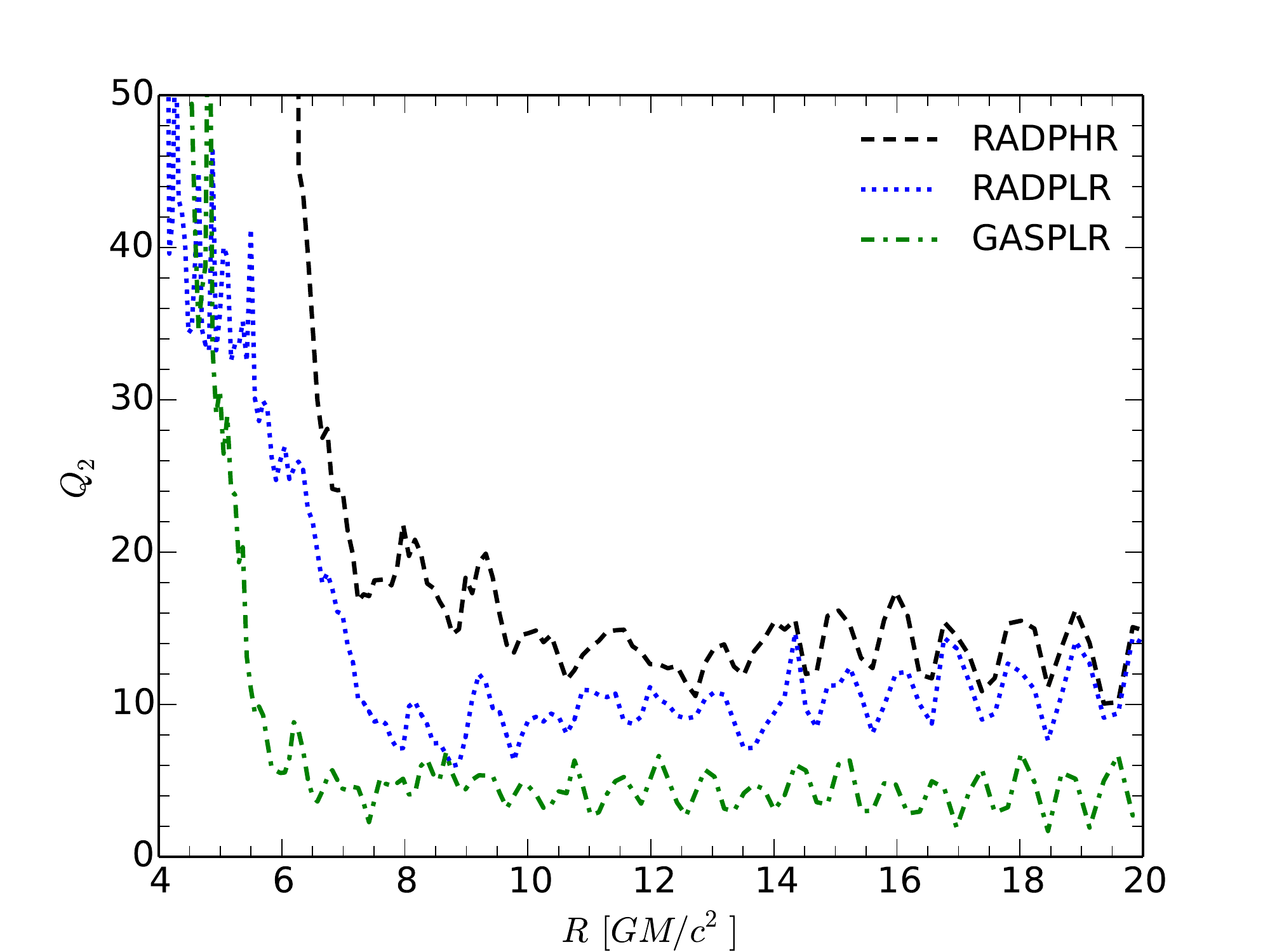}
\caption{Radial profiles of the vertical MRI Q parameter, $Q_2$, time averaged over the first $32\,t_\mathrm{ISCO}$ (the full duration of our GASPHR simulation).}
\label{mriq2}
\end{figure}

\begin{figure}
\centering
\includegraphics[width=1\columnwidth]{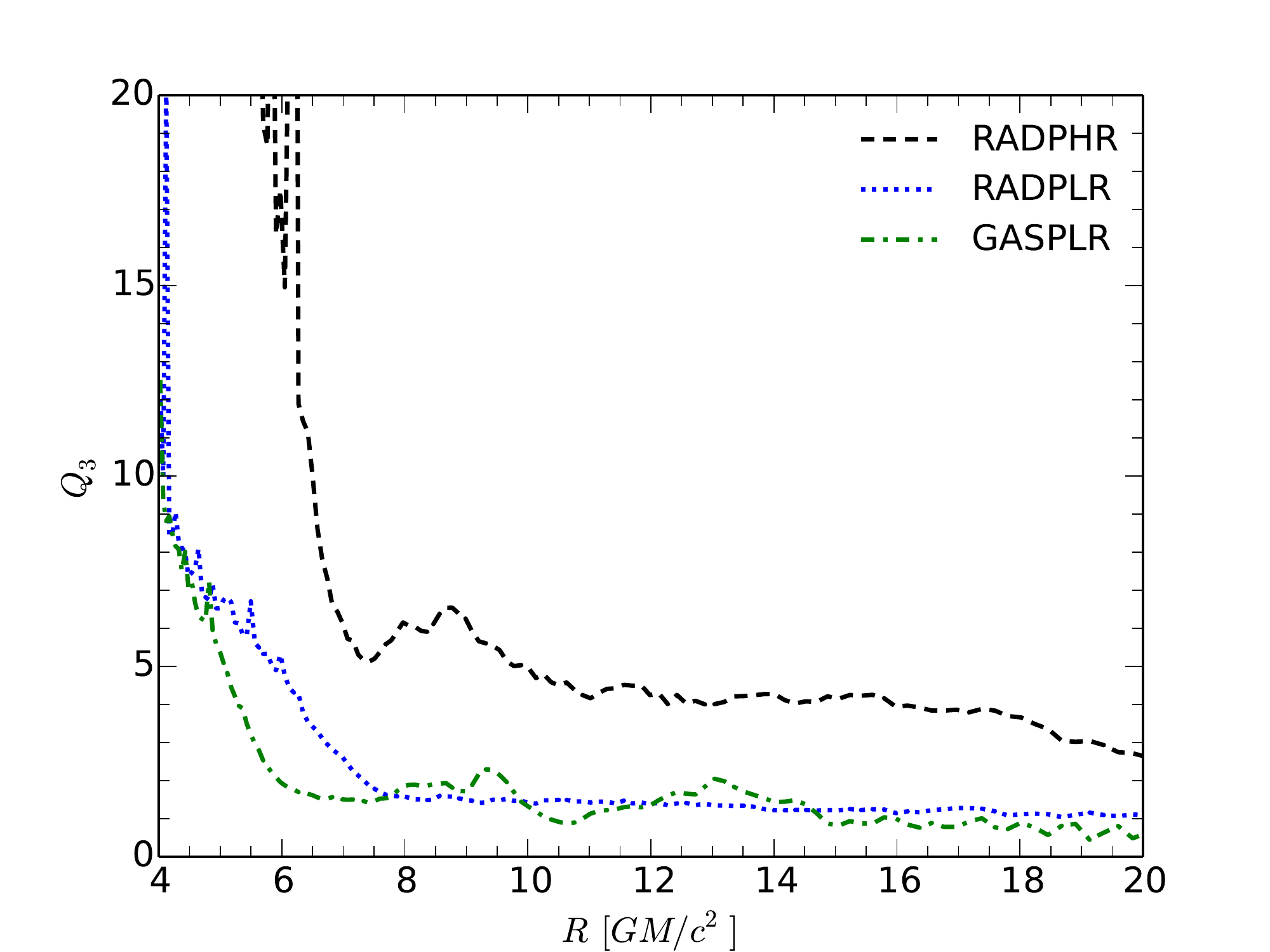}
\caption{Radial profiles of the azimuthal MRI Q parameter, $Q_3$, time averaged over the first $32\,t_\mathrm{ISCO}$.}
\label{mriq3}
\end{figure}

The standard measure of MRI resolution is the so-called $Q$ parameter, defined as 
\begin{equation}
Q_i = \frac{\lambda_{\mathrm{MRI},i}}{\Delta x_i}~,
\label{Q}
\end{equation}
where $\lambda_{\mathrm{MRI},i} = 2\pi v_{A,i}/|V^\phi|$ is the wavelength of fastest growing MRI mode, $\Delta x_i$ is a typical zone length, and $v_{A,i}=\sqrt{b^ib_i/\rho}$ is Alfven speed, all in a given direction, $i$. We checked both the vertical (Fig.~\ref{mriq2}) and azimuthal (Fig.~\ref{mriq3}) MRI $Q$ parameters for most of the models in Table.~\ref{models}. For our radiation-pressure-dominated simulations, we actually capture the vertical MRI well, with $Q_2 \gtrsim 10$ throughout most of the disc, especially in simulation RADPHR.  The gas-pressure-dominated simulation is not nearly as well resolved, with $Q_2 \lesssim 3$ at nearly all radii. The azimuthal direction is more problematic, with $Q_3 \lesssim 5$, even in the high-resolution RADPHR case. One way to possibly improve this in future simulations without requiring even more computational resources would be to reduce the azimuthal extent of the domain, while keeping the number of zones fixed. For now, we are left to point to the similarity of our results in both the low- and high-resolution simulations as evidence that the poor azimuthal MRI resolution does not negate our results of a thermal collapse in our radiation-pressure-dominated disc. 

\section{Discussion and Conclusions}

These are among the first global simulations of geometrically very thin discs in general relativity to include radiation \citep[see also][]{sadowski16}.  Our results are designed to be directly comparable to the shearing box simulations of \citet{jiang2013}, thus extending their results to global simulations. The major conclusions of our paper are:
\begin{enumerate}
\item As with previous shearing box \citep{jiang2013} and global simulations with weak magnetic fields \citep{sadowski16}, we find radiation-pressure-dominated thin discs to collapse.  In our RADPHR simulation, cooling dominated over heating throughout the disc, at least until late times. 
\item The strongest evidence for thermal instability is that the heating rate in the inner disc depends more strongly on mid-plane pressure than does the cooling rate (as shown in Figs.~\ref{coolpres} and  \ref{heatpres}), in agreement with standard theory.  Thus, any thermal equilibrium near the starting conditions of our RADP simulations would be unstable.
\item While the core of the radiation-pressure-dominated disc is collapsing, the position of the photosphere remains stable, as there is a magnetic-pressure-supported, optically thick atmosphere above the disc.
\item Our baseline gas-pressure-dominated simulation remained stable, with cooling and heating remaining roughly in balance throughout the disc. This confirms that our {\it Cosmos++} GRRMHD code is able to simulate stable, radiative, thin discs.  It also supports the conclusion that the collapse we found in the radiation-pressure-dominated case is not a numerical artifact.
\item The fact that the radiation-pressure-dominated disc collapses on roughly the local cooling time also suggests that the collapse is not due to numerical effects, such as under-resolved MRI, though we readily admit that there is room for improvement in this area of our simulations, particularly in the azimuthal direction. However, a comparison of our low- and high-resolution simulations suggests that our main conclusions are robust. 
\item We see evidence of one of the classic hallmarks of the viscous instability in the way our disc breaks up into rings.  To our knowledge, if confirmed this would be the first evidence of this instability in a numerical simulation.
\end{enumerate}

As with any numerical study, there are caveats to our results. We have only studied the stability of two disc configurations for relatively short evolution times.  A broader parameter study with longer simulations will be required to make our conclusions more robust.  It would be particularly good to consider two cases at a similar surface density, $\Sigma$, with one on the radiation-pressure-dominated (unstable) branch and the other on the gas-pressure-dominated (stable) branch. This will be addressed in a future paper.

\section{Acknowledgments}
The research was supported by Polish NCN grants 2013/08/A/ST9/00795 and 2014/15/N/ST9/04633. This research was also supported by the National Science Foundation under grant NSF AST-1211230. Simulations were done using PROMETHEUS supercomputer in the PL-Grid infrastructure. B.M. is also thankful to the College of Charleston for hosting him during the initial stages of this project. B.M. and P.C.F. thank the International Space Science Institute, where part of this work was carried out, for their hospitality.

\bibliographystyle{mn2e}
\bibliography{ref}

\begin{thebibliography}{23}
\expandafter\ifx\csname natexlab\endcsname\relax\def\natexlab#1{#1}\fi

\bibitem[{{Anninos}, {Fragile} \& {Salmonson}(2005){Anninos}, {Fragile}, \&
  {Salmonson}}]{anninos2005}
{Anninos} P., {Fragile} P.~C., {Salmonson} J.~D., 2005, \apj, 635, 723

\bibitem[{{Balbus} \& {Hawley}(1991)}]{balbus1991}
{Balbus} S.~A., {Hawley} J.~F., 1991, \apj, 376, 214

\bibitem[{{Begelman} \& {Pringle}(2007)}]{begelman07}
{Begelman} M.~C., {Pringle} J.~E., 2007, \mnras, 375, 1070

\bibitem[{{Brandenburg} {et~al}\mbox{.}(1995){Brandenburg}, {Nordlund},
  {Stein}, \& {Torkelsson}}]{bran1995}
{Brandenburg} A., {Nordlund} A., {Stein} R.~F., {Torkelsson} U., 1995, \apj,
  446, 741

\bibitem[{{Fragile} {et~al}\mbox{.}(2012){Fragile}, {Gillespie}, {Monahan},
  {Rodriguez}, \& {Anninos}}]{fragile2012}
{Fragile} P.~C., {Gillespie} A., {Monahan} T., {Rodriguez} M., {Anninos} P.,
  2012, \apjs, 201, 9

\bibitem[{{Fragile}, {Olejar} \& {Anninos}(2014){Fragile}, {Olejar}, \&
  {Anninos}}]{fragile2014}
{Fragile} P.~C., {Olejar} A., {Anninos} P., 2014, \apj, 796, 22

\bibitem[{{Hawley}, {Guan} \& {Krolik}(2011){Hawley}, {Guan}, \&
  {Krolik}}]{hawley2011}
{Hawley} J.~F., {Guan} X., {Krolik} J.~H., 2011, \apj, 738, 84

\bibitem[{{Hawley} {et~al}\mbox{.}(2013){Hawley}, {Richers}, {Guan}, \&
  {Krolik}}]{hawley2013}
{Hawley} J.~F., {Richers} S.~A., {Guan} X., {Krolik} J.~H., 2013, \apj, 772,
  102

\bibitem[{{Hirose}, {Krolik} \& {Blaes}(2009){Hirose}, {Krolik}, \&
  {Blaes}}]{hirose2009}
{Hirose} S., {Krolik} J.~H., {Blaes} O., 2009, \apj, 691, 16

\bibitem[{{Jiang}, {Stone} \& {Davis}(2013){Jiang}, {Stone}, \&
  {Davis}}]{jiang2013}
{Jiang} Y.-F., {Stone} J.~M., {Davis} S.~W., 2013, \apj, 778, 65

\bibitem[{{Lightman} \& {Eardley}(1974)}]{lightman1974}
{Lightman} A.~P., {Eardley} D.~M., 1974, \apjl, 187, L1

\bibitem[{{Mihalas} \& {Mihalas}(1984)}]{mihalas}
{Mihalas} D., {Mihalas} B.~W., 1984, {Foundations of radiation hydrodynamics}

\bibitem[{{Oda} {et~al}\mbox{.}(2009){Oda}, {Machida}, {Nakamura}, \&
  {Matsumoto}}]{oda09}
{Oda} H., {Machida} M., {Nakamura} K.~E., {Matsumoto} R., 2009, \apj, 697, 16

\bibitem[{{Piran}(1978)}]{piran1978}
{Piran} T., 1978, \apj, 221, 652

\bibitem[{{Pringle}(1976)}]{pringle1976}
{Pringle} J.~E., 1976, \mnras, 177, 65

\bibitem[{{Ressler} {et~al}\mbox{.}(2015){Ressler}, {Tchekhovskoy}, {Quataert},
  {Chandra}, \& {Gammie}}]{ressler2015}
{Ressler} S.~M., {Tchekhovskoy} A., {Quataert} E., {Chandra} M., {Gammie}
  C.~F., 2015, \mnras, 454, 1848

\bibitem[{{Reynolds} \& {Miller}(2009)}]{reynolds2009}
{Reynolds} C.~S., {Miller} M.~C., 2009, \apj, 692, 869

\bibitem[{{S{\c a}dowski}(2016)}]{sadowski16}
{S{\c a}dowski} A., 2016, \mnras, 459, 4397

\bibitem[{{S{\c a}dowski} {et~al}\mbox{.}(2013){S{\c a}dowski}, {Narayan},
  {Tchekhovskoy}, \& {Zhu}}]{2013MNRAS.429.3533S}
{S{\c a}dowski} A., {Narayan} R., {Tchekhovskoy} A., {Zhu} Y., 2013, \mnras,
  429, 3533

\bibitem[{{Shakura} \& {Sunyaev}(1973)}]{ss1973}
{Shakura} N.~I., {Sunyaev} R.~A., 1973, \aap, 24, 337

\bibitem[{{Shakura} \& {Sunyaev}(1976)}]{ss1976}
{Shakura} N.~I., {Sunyaev} R.~A., 1976, \mnras, 175, 613

\bibitem[{{Stone} {et~al}\mbox{.}(1996){Stone}, {Hawley}, {Gammie}, \&
  {Balbus}}]{stone1996}
{Stone} J.~M., {Hawley} J.~F., {Gammie} C.~F., {Balbus} S.~A., 1996, \apj, 463,
  656

\bibitem[{{Turner}(2004)}]{turner2004}
{Turner} N.~J., 2004, \apjl, 605, L45

\end{thebibliography}
\label{lastpage}
\end{document}